\documentclass[review]{elsarticle}

\usepackage{amsmath,amsthm,amsfonts,amssymb}
\usepackage{vhistory}
\usepackage{subfigure}

\usepackage{caption}
\usepackage[left=1.1in,right=1.1in,top=1in,bottom=1in]{geometry}

\newcommand{\Q}{{\mathbb{Q}}}

\newcommand{\E}{{\mathbb E}}

\newcommand{\1}{\mathbf{1}}

\newtheorem{definition}{Definition}[section]

\theoremstyle{definition} \newtheorem{remark}[definition]{Remark}
\theoremstyle{definition} \newtheorem*{remark*}{Remark}

\usepackage{algpseudocode}
\usepackage[ruled,vlined]{algorithm2e}

\begin{document}

\begin{frontmatter}

\title{Pricing and Risk Analysis in Hyperbolic Local Volatility Model with Quasi Monte Carlo}

\author{Julien Hok}
\address{Investec Bank, London, United Kingdom\\
		\it{julienhok@yahoo.fr}}

\author{Sergei Kucherenko} 
\address{BRODA, Ltd, London, United Kingdom\\
		\it{s.kucherenko@broda.co.uk}}

\begin{abstract}
Local volatility models usually capture the surface of implied volatilities more accurately
than other approaches, such as stochastic volatility models. We present the results of application
of Monte Carlo (MC) and Quasi Monte Carlo (QMC) methods for derivative pricing and risk analysis
based on Hyperbolic Local Volatility Model. In high-dimensional integration
QMC shows a superior performance over MC if the effective dimension of an integrand is not too large.
In application to derivative pricing and computation of Greeks effective dimensions
depend on path discretization algorithms. The results presented for the Asian option
show the superior performance of the Quasi Monte Carlo methods especially for the Brownian Bridge
discretization scheme.
\end{abstract}

\end{frontmatter}

\section{Introduction}

Monte Carlo (MC) methods are widely used in valuation of complex
financial instruments. Although the convergence rate of MC methods is $O(\frac{1}{\sqrt{N}})$
,where $N$ is the number of sampled points does not depend
on the number of variables $n$ but it is rather slow.
Switching from random numbers to quasi-random numbers such as
low-discrepancy sequences (LDS) can significantly improve
the convergence under some conditions. Methods based on LDS are known
as Quasi Monte Carlo (QMC). Asymptotically, they can provide a rate of
convergence $O(\frac{1}{N})$. Quality of Sobol' sequences heavily depends on
the so-called direction numbers and in practice very few Sobol'
sequence generators show good efficiency in valuation of complex
financial instruments (see f.e. \cite{BRODA, SobAsoKreiKuch11}).

The solution of financial problem such as pricing and hedging can be formulated as a mathematical expectation of some functionals,
which in practice is reduced to evaluation of the Wiener path integrals. The nominal dimensions
of such integrals is the product of the number of time steps at which the asset prices are observed
and the number of risk factors (the underlying assets). It can reach many thousands of dimensions.
QMC methods can loose superior performance in high-dimensional settings unless problem's effective dimension is low.
The concept of effective dimensions was introduced in \cite{CaflischMorokoffOwen97}. It was shown in many papers that QMC is superior to MC if the effective dimension of
an integrand is not too large. Effective dimensions and the QMC convergence of path dependent integrals
depend on path discretization algorithms.
There are two widely used in finance algorithms. They are 1) the incremental (aslo known as standard) discretization algorithm
2) the Brownian bridge discretization algorithm. Both algorithms have the same variance,
hence their MC convergence rates are the same. However, the corresponding QMC algorithms
have different efficiencies with the Brownian bridge having much higher convergence rate for majority of payoffs.
This combination of Sobol points with the Brownian bridge construction was proposed
by Moskowitz and Caflisch \cite{MoskowitzCaflisch1996} and has been found to be highly effective in finance applications
(see e.g \cite{AcworthBroadieGlasserman98,AkessonLehoczky00,BianchettiKucherenkoScoleri15, KucherenkoShah07,CaflischMorokoffOwen97}).

Pricing and hedging of financial instruments has been primarily based on {\it{Gaussian}} models,  where the underlying asset
dynamics (interest rates, equity pricess or exchange rates) are assumed to follow a {\it{Hull-White}}  or {\it{Black-Scholes}} models.
However, empirical asset returns distributions tend to exhibit fat-tails (kurtosis) and skewness (asymmetric distribution).
The {\it{skew}} or {\it{smile}} in {\it{implied volatility}} surfaces (defined in Section \ref{section:HLV Model}) observed across various asset classes are market reality (see e.g \cite{Gatheral06,OverhausAnaHans07,Wilmott06}).
We need more convenient models for the asset $S$ able to produce more closely the implied volatility surfaces.
Local volatility models, either parametric or non-parametric (see e.g. \cite{Dupire94,DerKa98,Rubi94,Jac08} or \cite{Cox75}),
usually capture the surface of implied volatilities more accurately than other approaches, such as stochastic volatility models ( see e.g. \cite{MadQianRen07,Romo12} for discussions). Here we extend previous analysis to a more realistic local volatility type diffusion, namely the {\it{hyperbolic local volatility}} introduced in \cite{Jac08} and widely used in quantitative finance industry (see e.g \cite{BompisHok14,HokNGareAntonis18,HokShih2019}).

The objective of this paper is to compare application of MC and QMC methods for pricing financial derivatives and computation of Greeks
using hyperbolic local volatility model. The rest of this paper is organized as follows.
Section 2 briefly reviews MC and QMC methods. Section 3 introduces
time-homogeneous hyperbolic local volatility model. Time discretization schemes are
presented in Section 4. MC simulation of option pricing and computation of Greeks
are considered in Section 5. Section 6 presents the results of
prices and sensitivities (Greeks) computation.
Finally, conclusions and directions of future work are given in Section 6.

\section{MC and QMC algorithms} \label{section:MCandQMCalgorithms}

Option pricing problem after transformation can be reduced to the computation of the multidimensional  integral

\begin{equation}\label{multiDintegral}
I[f] = \int_{H^n} f(x)dx.
\end{equation}
Here function $f(x)$ is integrable in the $n$-dimensional unit hypercube $H^n$.
The justification is given in Section \ref{sect:mc_option_pricing}.
The MC quadrature formula is based on the probabilistic interpretation
of an integral. For a random variable that is uniformly distributed in $H^n$

\begin{equation}\label{multiDintegralProbaInterp}
I[f] = \E [f(x)],
\end{equation}
where $\E[f(x)]$ is the mathematical expectation. The standard MC estimator of an
expectation is

\begin{equation}\label{multiDintProbaInterpProxy}
I_N[f] = \frac{1}{N} \sum_{i=1}^N f(x_i),
\end{equation}
where $\{x_i\}$ is a sequence of random points in $H^n$ of length $N$. The approximation
$I_N[f]$ converges to $I[f]$ with probability $1$. An integration error
$\epsilon$ according to the Central Limit Theorem has the expectation

\begin{equation}\label{MCConvSpeed}
\E(\epsilon^2) =  \frac{\sigma^2(f)}{N},
\end{equation}
where $\sigma^2(f)$ is the function variance. Then the expression for
the root mean square error of the MC method is

\begin{equation}\label{MCMeanSqError}
\epsilon_N = (\E(\epsilon^2))^{\frac{1}{2}} = \frac{\sigma(f)}{\sqrt{N}}.
\end{equation}

The convergence rate of MC does not depend on the number of variables
$n$ but it is rather slow. It is known that random number sampling is prone to clustering.
As new points are added randomly, they do not necessarily fill the gaps between already sampled points.
On the other hand QMC methods are based on LDS
(also known as {\it{quasi random numbers}}). LDS are specifically designed to place sample points as uniformly as possible.
Successive LDS points “know” about the position of previously sampled points and “fill” the gaps between
them. The QMC algorithm
for the evaluation of the integral (\ref{multiDintegralProbaInterp}) has a form similar to (\ref{multiDintProbaInterpProxy})
where instead of a sequence of random points $\{x_i\}$ LDS points $\{q_i\}$ uniformly
distributed in a unit hypercube $H^n$ are used: $q_i = (q_i^1,...,q_i^n)$ .

Many practical studies have proven that the Sobol LDS \cite{Sobol67} is in many aspects superior to other LDS (see e.g \cite{Jackel02,Wilmott06,KucherenkoShah07,BianchettiKucherenkoScoleri15,Glasserman03}).
For this reason it was used in this work. Sobol LDS were constructed by following the three main requirements (\cite{Sobol67}):

\begin{itemize}
\item Best uniformity of distribution as $N \to +\infty$;
\item Good distribution for fairly small initial sets;
\item A very fast computational algorithm.
\end{itemize}

Points generated by the Sobol LDS produce a very uniform filling of the
space even for a rather small number of points $N$, which is a very important
case in practice.

In this work we used Sobol sequence generator \texttt{SobolSeq}
generator provided by BRODA Ltd. \cite{BRODA}\footnote{The most recent generator released by BRODA is \texttt{SobolSeq65536} with maximum dimensionality 65536 and Properties A and A'.}.
Sobol' sequences produced by \texttt{SobolSeq} satisfy additional uniformity properties: Property A
for all dimensions and Property A' for adjacent dimensions (see
\cite{SobAsoKreiKuch11} for details). It has been
found that BRODA's \texttt{SobolSeq} generators
outperforms all other known LDS generators both in speed and
accuracy \cite{Renz2018, SobAsoKreiKuch11}.

For the best known LDS sequences the estimate for the rate of convergence
$I_N \to I$ is known to be $\frac{O(ln^n N)}{N}$, while for Sobol' LDS $\frac{O(ln^{(n-1)} N)}{N}$ if $N=2^p$, where $p$ is an integer.
This rate of convergence is much faster than that for the MC method (\ref{MCMeanSqError}),
although it depends on the dimensionality $n$. Consequently, the smaller $n$, the better this estimate. In practice
at $n > 1$ the rate of convergence $\frac{O(ln^n N)}{N}$ is not observed. It appears
to be approximately $N^{- \alpha} , 0 < \alpha \leq 1$ depending on the effective dimension.
For financial problems typically $0.5 < \alpha \leq 1$. Hence, the QMC method usually outperforms MC in terms of convergence.
$\alpha$ can be dramatically increased by using effective dimension reduction techniques
such as the {\it{Brownian Bridge}}.


\section{Time-homogeneous hyperbolic local volatility model} \label{section:HLV Model}

\begin{figure}[htbp]
  \centering
 \includegraphics[scale=0.6, angle =90,trim={0.5cm 3.5cm 1cm 2.5cm}, clip]{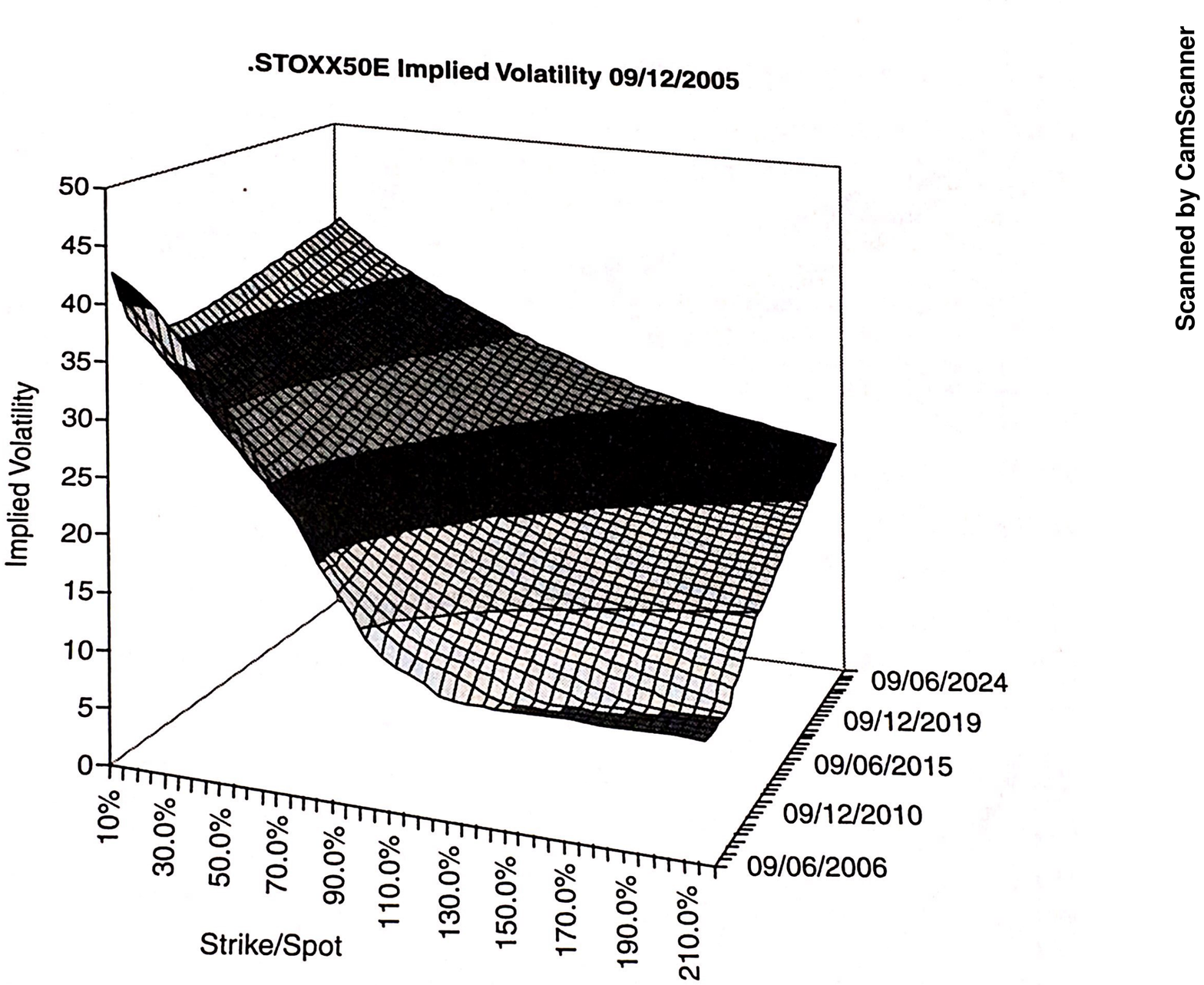}
  \caption{Implied volatilities for different strikes and maturities for STOXX50E on the 09/12/2005.}
  \label{figure:ImpliedVolEUROSTOXX50}
\end{figure}

Since the advent of the Black-Scholes option pricing formula, the study of implied volatility has become a central preoccupation for both academics and practitioners. It is well known, actual option prices rarely conform to the predictions of explicit formulas because the idealized assumptions required for it to hold don't apply in the real world. Consequently, implied volatility (the volatility input to the Black-Scholes formula that generates the market European Call or Put price) in general depends on the strike $K$ and the maturity of the option $T$. The collection of all such implied volatilities
is known as the volatility surface. The effect that implied volatility $\sigma_{im}(T,K)$
is a decreasing function of strike is called {\it{skew}}. Figure \ref{figure:ImpliedVolEUROSTOXX50} provides an illustration for the equity index {\it{STOXX50E}}.
The graph shows a strong {\it{skew}} for all maturities and this shape is usually observed in the equity derivatives market. This means that the underlying asset price process cannot be explained using the Black-Scholes model, for which the implied volatility does not depend on the strike.
Rather, we need to find a convenient model for the underlying asset to evaluate contingent claims.
Local volatility models, either parametric or non-parametric, see e.g \cite{Dupire94,DerKa98,Rubi94,Jac08} or \cite{Cox75}, usually capture the surface of implied volatilities more precisely than other approaches, such as stochastic volatility models (see {\cite{MadQianRen07, Romo12}} for details).

For our analysis, we consider the time homogeneous
hyperbolic local volatility model (HLV) which is widely used in quantitative finance to
capture the market skew. It corresponds to a parametric local volatility-type model in which the dynamic of the underlying
under risk neutral measure $\Q$ is:

\begin{align} \label{HyperbolicSDE1}
d S(t) = rS(t) dt + \tilde{\sigma} (S(t)) dW(t), \ S_0=1,
\end{align}

where $r$ is the risk free interest rate and

\begin{equation}
\tilde{\sigma} (S) = \nu \Big\{ \frac{(1-\beta+\beta^2)}{\beta} S +\frac{(\beta-1)}{\beta}  \big(\sqrt{S^2+\beta^2(1-S)^2}-\beta\big) \Big\}
\end{equation}

\begin{figure}[htbp]
  \centering
 \includegraphics[scale=0.6, trim={0.5cm 7cm 1cm 6cm}, clip]{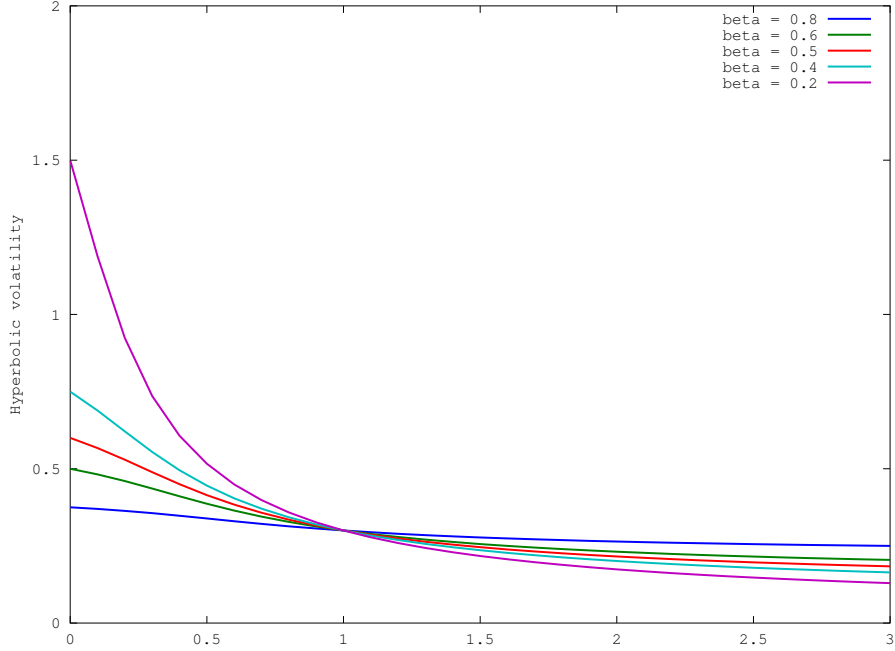}
  \caption{ Impact of the value $\beta$ on the hyperbolic local volatility for fixed volatility level $\nu = 0.3$. }
  \label{figure:HyperbolicLVImpactBeta}
\end{figure}

Here $\nu >0$ is the level of volatility, $\beta \in (0,1]$ is the skew
parameter and $W$ is the standard Brownian motion. This model was introduced in \cite{Jackel10}.
It behaves similarly to the Constant Elasticity of Variance (CEV) model, and has been used for numerical experiments in
\cite{BompisHok14,HokNGareAntonis18,HokShih2019}.
The advantage of this model is that zero is not an attainable boundary, and that allows to avoid
some numerical instabilities present in the CEV model when the underlying asset price
is close to zero (see e.g. \cite{And00}).
It corresponds to the Black-Scholes model for $\beta=1$ and exhibits a skew for the implied volatility surface when $\beta \neq 1$. Figure (\ref{figure:HyperbolicLVImpactBeta}) illustrates the
impact of the parameter $\beta$ on the skew of the volatility surface. We observe that the skew increases significantly with decreasing value of $\beta$. For example with $\nu = 0.3, \, \beta = 0.2$, the difference in volatility between strikes at $50\%$ and at $100\%$ is about $15 \%$.\\

\section{Time discretization schemes} \label{sect:time discretization schemes}

\subsection{Euler discretization of the SDE} \label{subsection:euler discretization}

Consider the problem of pricing an option on a single asset whose value at time $t$ is denoted by $S(t)$.
We assume the asset follows a HLV process defined by SDE (\ref{HyperbolicSDE1}). To guarantee positive price in the simulation,
let's pose $Y(t) = \ln (S(t))$ and by Ito formula, we have

\begin{align} \label{HyperbolicSDEY}
d Y(t) = [r -\frac{1}{2} \sigma^2(Y(t))] dt + \sigma(Y(t))     dW_t, \ Y(0)= \log (S(0)),
\end{align}
with $\sigma(Y) = \frac{\tilde{\sigma}(e^Y)}{e^Y}$.\\

As the solution is not known in closed form, we proceed by using the discretisation Euler-Maruyama scheme
(\cite{Glasserman03,Kloeden:Platen91,Maruyama55}).
In a discrete case of $n$ equally distributed time steps,  it has following form:

\begin{align} \label{HyperbolicSDEXEuler}
Y^{n}(t_{i+1}) = Y^{n}(t_{i}) +  [r-\frac{1}{2} \sigma^2(Y^{n}(t_{i}))] (t_{i+1} - t_{i}) +
\sigma(Y^{n}(t_{i})) \sqrt{t_{i+1} - t_{i}}(W(t_{i+1}) - W(t_{i}))
\end{align}
with $Y^{n}(0) = \log (S(0))$, $ \Delta t = \frac{T}{n}, t_i = i \Delta t, \, i=0,..,n$.\\

In addition to the statistical noise discussed in Section \ref{section:MCandQMCalgorithms}, there is also a discretisation error associated to a chosen discretisation scheme.
Theorem 10.2.2 in \cite{Kloeden:Platen91} provides conditions for Euler-Maruyama scheme to have a strong error
convergence of order $\frac{1}{2}$ . Under stronger conditions as in \cite{Kloeden:Platen91}, theorem 14.5.2, the scheme reaches a weak error convergence of order 1.

\subsection{Discretization of the Wiener process}

We consider two algorithms for the discretization of the Brownian motion $W$ in equation (\ref{HyperbolicSDEY}). The first one is known as the incremental (standard)  discretization algorithm. Its construction follows directly from the definition of $W(t)$. The second one is the alternative
discretization algorithm which is based on the use of conditional distributions.

The standard (incremental) discretization algorithm is defined by the relation:

\begin{equation}\label{StdDis}
W(t_i) = W(t_{i-1}) + \sqrt{\Delta t} Z_i \,\,\, 1 \leq i \leq n,
\end{equation}
where $(Z_i)$ are independent standard normal variates. In the standard
discretization algorithm the evolution of an asset value is
generated by normal variates with equal weights.

In the Brownian bridge discretization, the value of $W(t_i)$ is generated
from values of $W(t_l),W(t_m), l \leq i \leq m$ at earlier and later time steps.
Unlike the standard discretization, which generates $W(t_{i+1})$ sequentially
along the time horizon, the Brownian bridge discretization first generates
the variable at the terminal point

$$W(T) = \sqrt{T}Z_1$$
and then it fills other points using already found values of $W(t_i)$. The generalised Brownian bridge formula is given by
\begin{equation} \label{BBFormula}
W(t_i) = (1 - \gamma)W(t_l) + \gamma W(t_m) + \sqrt{\gamma(1 - \gamma)(m - l) \Delta t}Z_i,
\end{equation}
where $\gamma = \frac{i-l}{m-l}$ (\cite{Morokoff98}). It can be seen from equation (\ref{BBFormula}) that the variance of the stochastic part of the Brownian bridge formula
$\gamma(1 - \gamma)(m - l) \Delta t$. It decreases at the successive levels of refinement
and the first few points contain the most of the variance.
This variance is less than that in (\ref{StdDis}) as $\gamma (1 - \gamma)(m - l) < 1$.
Both algorithms have the same variance, hence their MC convergence rates
are the same. However, QMC algorithms have different efficiencies with the
Brownian bridge algorithm having a much higher convergence rate
(see e.g \cite{CaflischMorokoffOwen97,SobolKucherenko05,KucherenkoShah07,BianchettiKucherenkoScoleri15}).

\remark \label{GaussianUniformTransform}{
Standard normal variates are computed as $Z_i \overset{L}{=} \Phi^{-1}(U_i)$, where $\Phi$
is the cumulative function of normal distribution and $U_i$ is a random variable with uniform distribution in $[0,1]$.
So in practice, one simulates independent uniform random variable and use this transformation to obtain independent standard Gaussian variables.}

\section{Monte Carlo simulation of option pricing and computation of Greeks} \label{sect:mc_option_pricing}

\subsubsection{Option pricing} \label{subsect:option_pricing}

We consider a geometric average Asian call option whose payoff function is given by

\begin{equation} \label{asianpayoff}
P_A = \max (\bar{S} - K, 0),
\end{equation}
where $\bar{S}$ is a geometric average at $n$ equally spaced time point:

\begin{equation} \label{geometricaverage}
\bar{S} = ( \prod_{i=1}^n S_i)^{\frac{1}{n}},
\end{equation}
where $S_i$ is the asset price at time $t_i = i \frac{T}{n}$, $1 \leq i \leq n$.

In a risk neutral environment, the value of a geometric average Asian call option with maturity $T$ and strike $K$ is the discounted
value of its payoff:

\begin{equation} \label{priceasianpayoff}
AC(T,K) = e^{-rT} \E^{\Q}[P_A].
\end{equation}

In the HLV model, there is no analytical formula for (\ref{priceasianpayoff}). We are going to estimate the price by the MC method. There are two steps:
Firstly, we approximate the asset price $S(t_i)$ with $S^n(t_i) = e^{Y^n(t_i)}$ by discretising
the SDE (\ref{HyperbolicSDEY}) as described in Section \ref{subsection:euler discretization};
Secondly, the MC method approximates the expectation of the Asian payoff (\ref{asianpayoff})
with a simple arithmetic average of payoffs taken over a finite number $N$ of simulated price paths:

\begin{equation} \label{MCpriceasianpayoff}
AC_N(T,K) = e^{-rT} \left[ \frac{1}{N} \sum_{i=1}^N \max (\bar{S}^{(i)} - K, 0) \right].
\end{equation}
where $\bar{S}^{(i)}$ is an approximation of $\bar{S}$ using the simulated price paths $i$.
So $e^{-rT} \max (\bar{S}^{(i)} - K, 0)$ can be written as $f(U_{i1}, U_{i2},.., U_{in})$ following the remark \ref{GaussianUniformTransform} where all $(U_{ij})$ are independent uniform variates, which
together with (\ref{priceasianpayoff})
justifies formula (\ref{multiDintegral}).

\subsubsection{Sensitivity factors}

Sensitiviry factors or {\it{Greeks}} are derivatives of the price $AC(T,K)$ w.r.t specific parameters like spot price or volatility. They are very important quantities which need to be computed for hedging and risk management purposes. In the present work, we consider in particular the following Greeks:

\begin{align} \label{Greeks}
\Delta &= \frac{\partial AC(T,K) }{\partial S(0)}, \\
\Gamma &= \frac{\partial^2 AC(T,K) }{\partial S(0)^2}, \\
\vartheta_{\nu} &= \frac{\partial AC(T,K) }{\partial \nu}, \\
\vartheta_{\beta} &= \frac{\partial AC(T,K) }{\partial \beta}
\end{align}
called Delta, Gamma, $\nu$-Vega and $\beta$-Vega respectively. $Delta$ represents the hedge of the financial instrument w.r.t the risky underlying $S$. In the {\it{dynamic hedging}}, it corresponds to the number of assets hold that must be continuously changed to maintain a {\it{delta-neutral}} position. Gamma is the second derivative of the price with respect
to the underlying. Since gamma is the sensitivity of the delta to the underlying it is a measure of how much or how often a position must be rehedged in order to maintain a delta-neutral position.
The vega represents the sensitivity of the option price to volatility. As discussed in Section
\ref{section:HLV Model}, $\nu$-Vega measures the price sensitivity to the level of volatility and
$\beta$-Vega to the volatility skewness.
As there are no analytical formulas, the Greeks above are estimated by MC simulation and finite differences method, using the central difference formulas:

\begin{align} \label{GreeksEstimation}
\Delta & \approx \frac{ AC_N(T,K, S(0) + \epsilon_s) - AC_N(T,K, S(0) - \epsilon_s) }{2 \epsilon_s}, \\
\Gamma &\approx \frac{ AC_N(T,K, S(0) + \epsilon_s) + AC_N(T,K, S(0) - \epsilon_s)
-2 AC_N(T,K, S(0))
}{\epsilon_s^2}, \\
\vartheta_{\nu} & \approx \frac{ AC_N(T,K, \nu + \epsilon_v) - AC_N(T,K, \nu - \epsilon_v) }{2 \epsilon_v} , \\
\vartheta_{\beta} & \approx \frac{ AC_N(T,K, \beta + \epsilon_v) - AC_N(T,K, \beta - \epsilon_v) }{2 \epsilon_v}.
\end{align}

In the MC simulations for Greeks we use path recycling of both pseudo-random sequences and LDS to
minimize the variance of the Greeks, as suggested in \cite{Jackel02,Glasserman03}.
Notice that the analysis of the RMSE for Greeks is, in general, more complex
than that for prices, since the variance of the MC simulation mixes with the bias due
to the approximation of derivatives with finite differences. For the sensitivity factors estimation to be meaningful and not entirely hidden by MC noise, the shift $\epsilon_s$ and $\epsilon_v$ are chosen to be large enough and represent about $1 \%$ of the current spot and the volatility parameters  $\beta$ respectively (see \cite{Jackel02,Glasserman03} for detailed discussions).

\section{Numerical results}

In this Section we present the numerical results from simulations of
prices and sensitivity factors for Asian call options. The following parameters were used for simulation: $S_0=100, \, r = 3 \%, \, T = 1, \ \nu = 30\%, \, \beta = 0.5$,
number of discrete time steps $n$ = 256. It corresponds to a high dimension case.
For thoroughness, we consider {\it{in-the-money}}, {\it{at-the-money}} and {\it{out-the-money}} options with strike $80, \, 100, \, \120$ respectively.

There are no analytical solutions for geometric average Asian call option prices and sensitivity factors in the HLV model.
Numerical simulations using MC and QMC methods were performed
to compare convergence of the methods with the standard and the Brownian Bridge discretization schemes. The Mersenne Twister generator,
which is considered to be one of the most efficient uniform random
number generators was used for MC simulation (\cite{MatsumotoNishimura98}). The Sobol sequence generator {\it{SobolSeq}} with additional uniformity properties
described in Section 2 was used for QMC simulations
(\cite{BRODA,SobAsoKreiKuch11}). To eliminate randomness which may be caused by the seed point for MC or a specific set of Sobol' points, we performed $L=10$ independent runs.
For the MC method all runs were statistically independent. For QMC integration for each run a different part of the Sobol sequence was used. We denote by $Q_N^{(l)}$ a quantity (price or sensitivity factors) on $l-$th run for $N$ paths replications. Each estimated quantity
$\bar{Q}_N$ was averaged over $L=10$ independent runs i.e

\begin{equation}
\bar{Q}_N = \frac{1}{L} \sum_{l=1}^L Q_N^{(l)}.
\end{equation}

The reference values were estimated by the Sobol QMC method with $m = 262,144$ simulation paths and given by

\begin{equation}
\bar{Q}_{ref} = \frac{1}{L} \sum_{l=1}^L Q_m^{(l)}.
\end{equation}

\begin{figure}[!tbp]
  \centering
\subfigure[]
{
  \begin{minipage}[b]{5.5cm}
    \includegraphics[width=8cm, height=4cm, angle =0,trim={2cm 18.5cm 1cm 2.5cm}, clip]{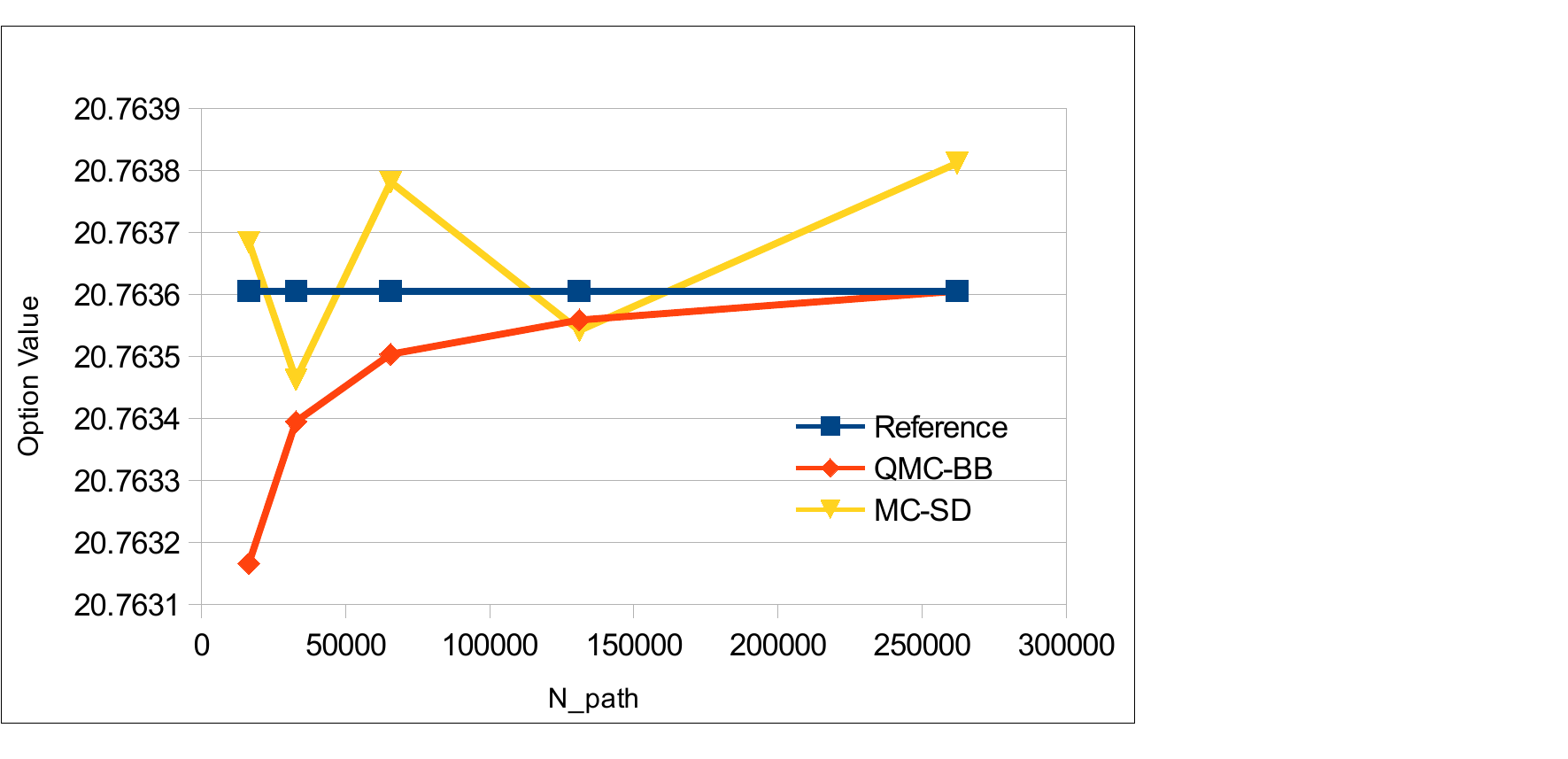}
    \label{figure:OptionValueK80}
  \end{minipage}
  }
  \hfill
  \subfigure[]
  {
  \begin{minipage}[b]{5.5cm}
   \includegraphics[width=8cm, height=4cm, angle =0,trim={2cm 18.5cm 1cm 2.5cm}, clip]{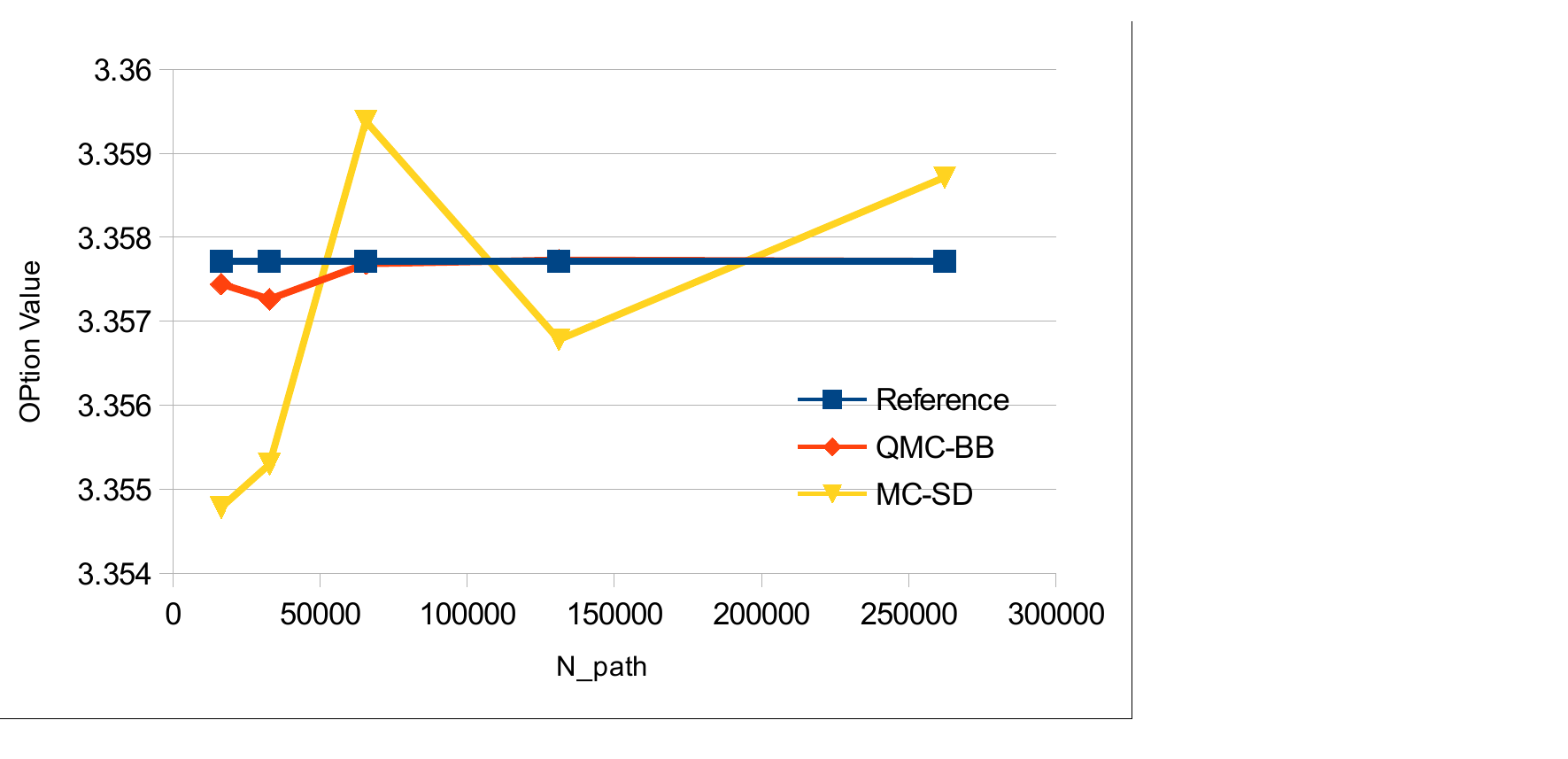}
    \label{figure:OptionValueK100}
  \end{minipage}
  }
  \subfigure[]
  {
   \begin{minipage}[b]{5.5cm}
    \includegraphics[width=8cm, height=4cm, angle =0,trim={2cm 18.5cm 1cm 2.4cm}, clip]{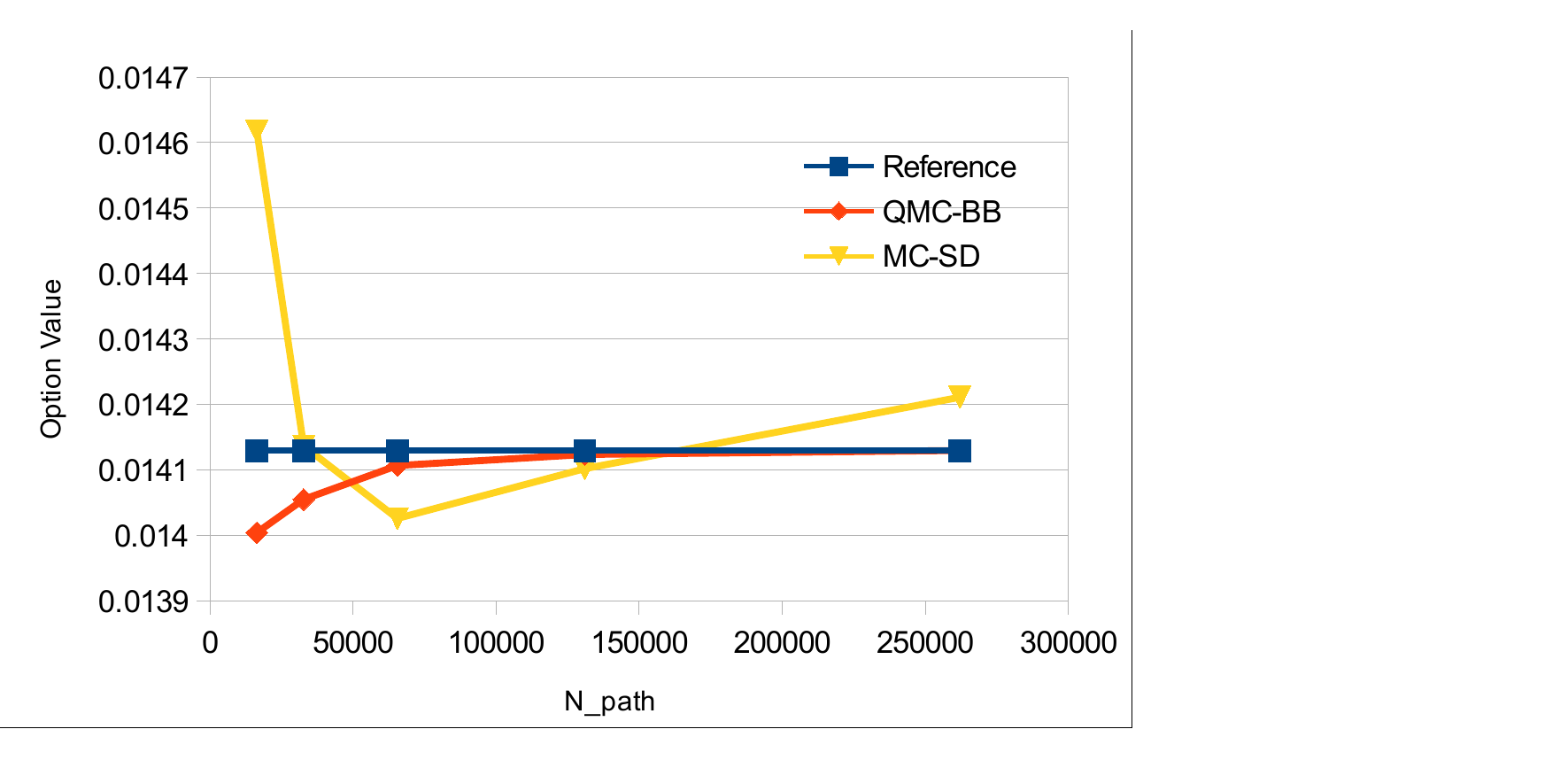}
	\label{figure:OptionValueK120}
  \end{minipage}
  }
\caption{Asian call price with strike (a) 80; (b) 100; (c) 120.}
\label{figure:OptionValueK80-120}
\end{figure}

\begin{figure}[!tbp]
  \centering
  \subfigure[]
{
  \begin{minipage}[b]{5.5cm}
    \includegraphics[width=8cm, height=4cm, angle =0,trim={2cm 18.5cm 1cm 2.5cm}, clip]{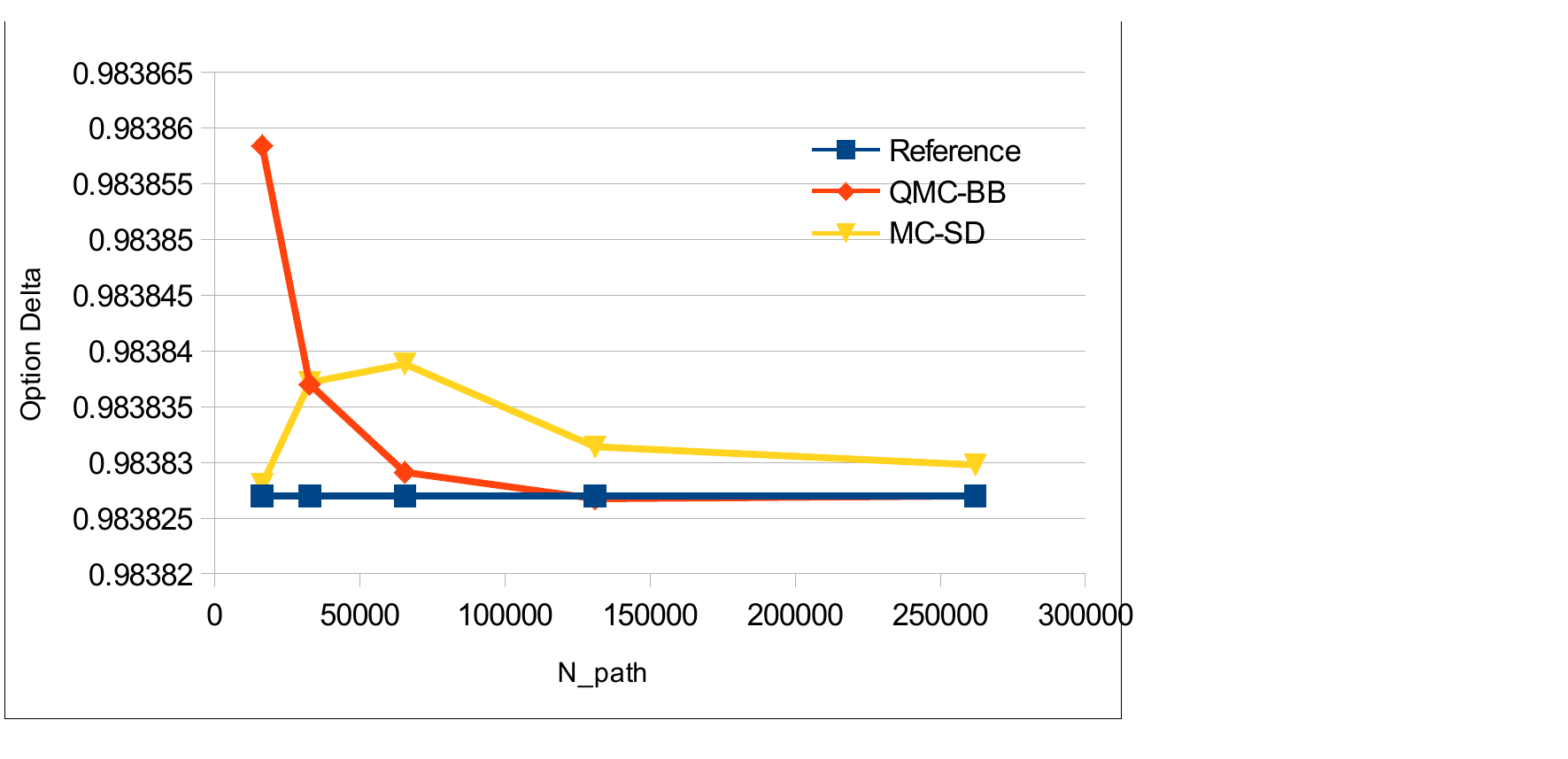}
    \label{figure:OptionDeltaK80}
  \end{minipage}
  }
  \hfill
  \subfigure[]
{
  \begin{minipage}[b]{5.5cm}
    \includegraphics[width=8cm, height=4cm, angle =0,trim={2cm 18.5cm 1cm 2.5cm}, clip]{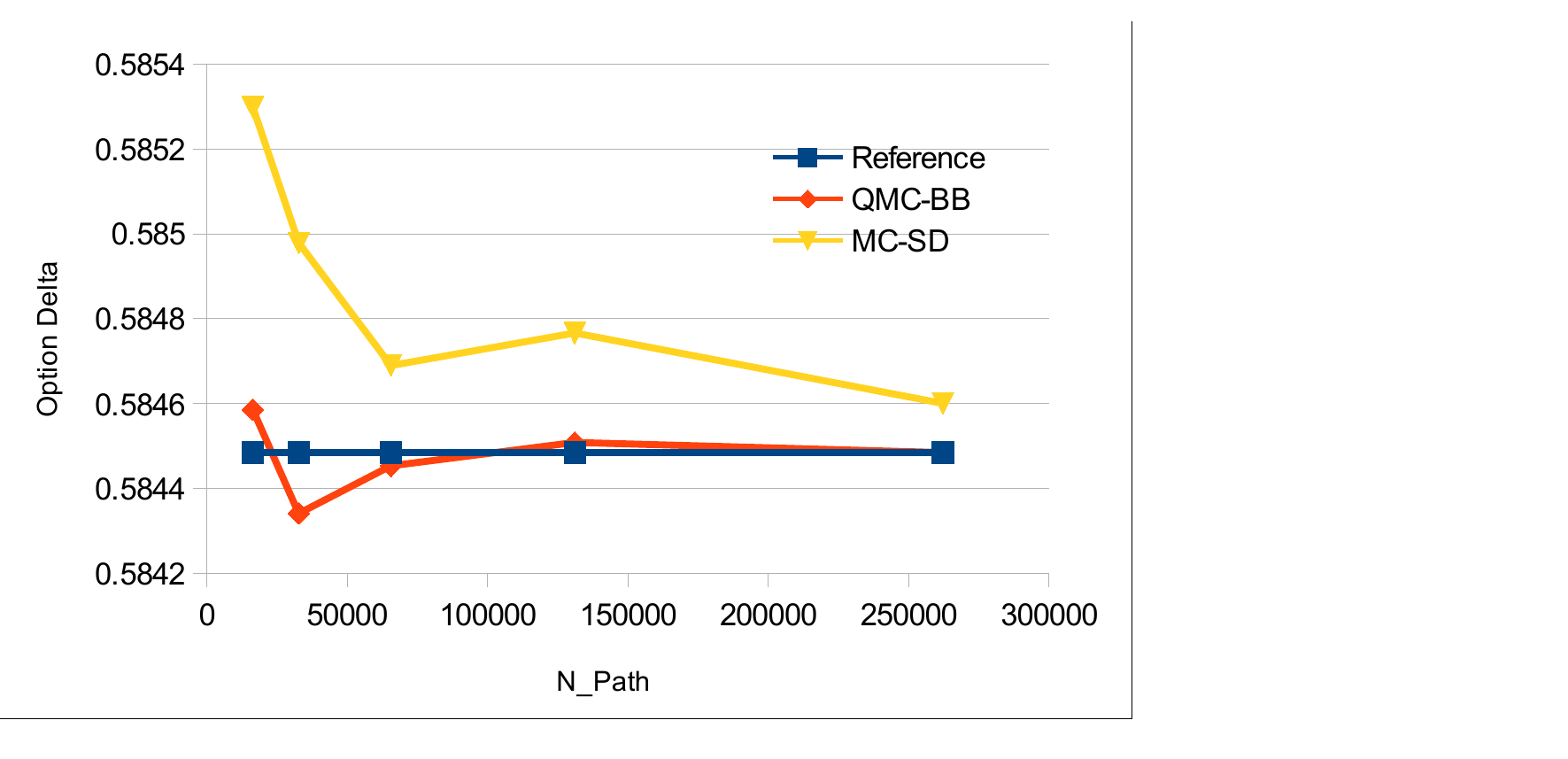}
    \label{figure:OptionDeltaK100}
  \end{minipage}
  }
  \subfigure[]
{
   \begin{minipage}[b]{5.5cm}
    \includegraphics[width=8cm, height=4cm, angle =0,trim={2cm 18.5cm 1cm 2.4cm}, clip]{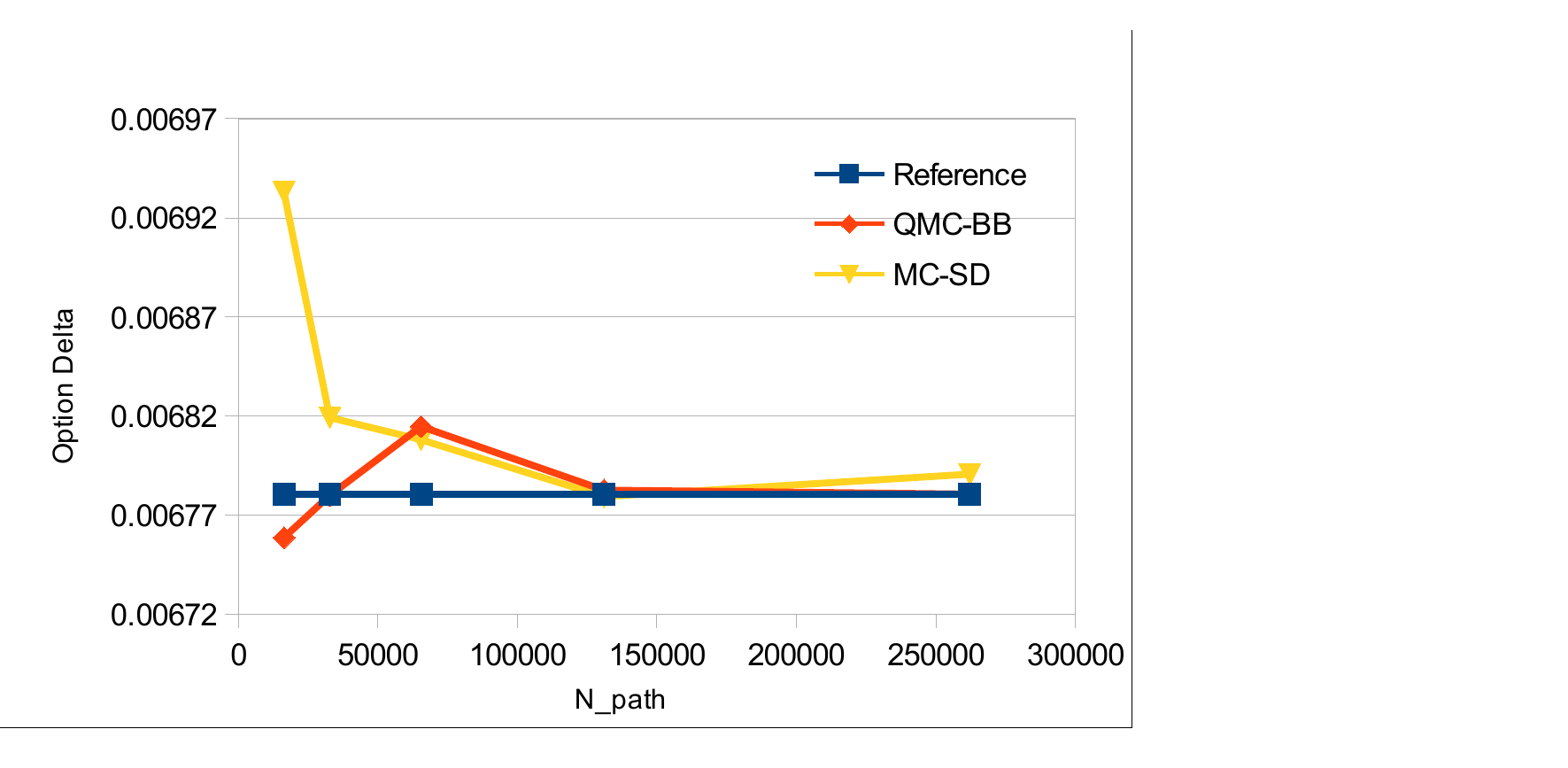}
		\label{figure:OptionDeltaK120}
  \end{minipage}
  }
\caption{Asian call delta value with strike (a) 80; (b) 100; (c) 120.}
\label{figure:OptionDeltaK80-120}
\end{figure}

\begin{figure}[!tbp]
  \centering
   \subfigure[]
{
  \begin{minipage}[b]{5.5cm}
    \includegraphics[width=8cm, height=4cm, angle =0,trim={2cm 18.5cm 1cm 2.5cm}, clip]{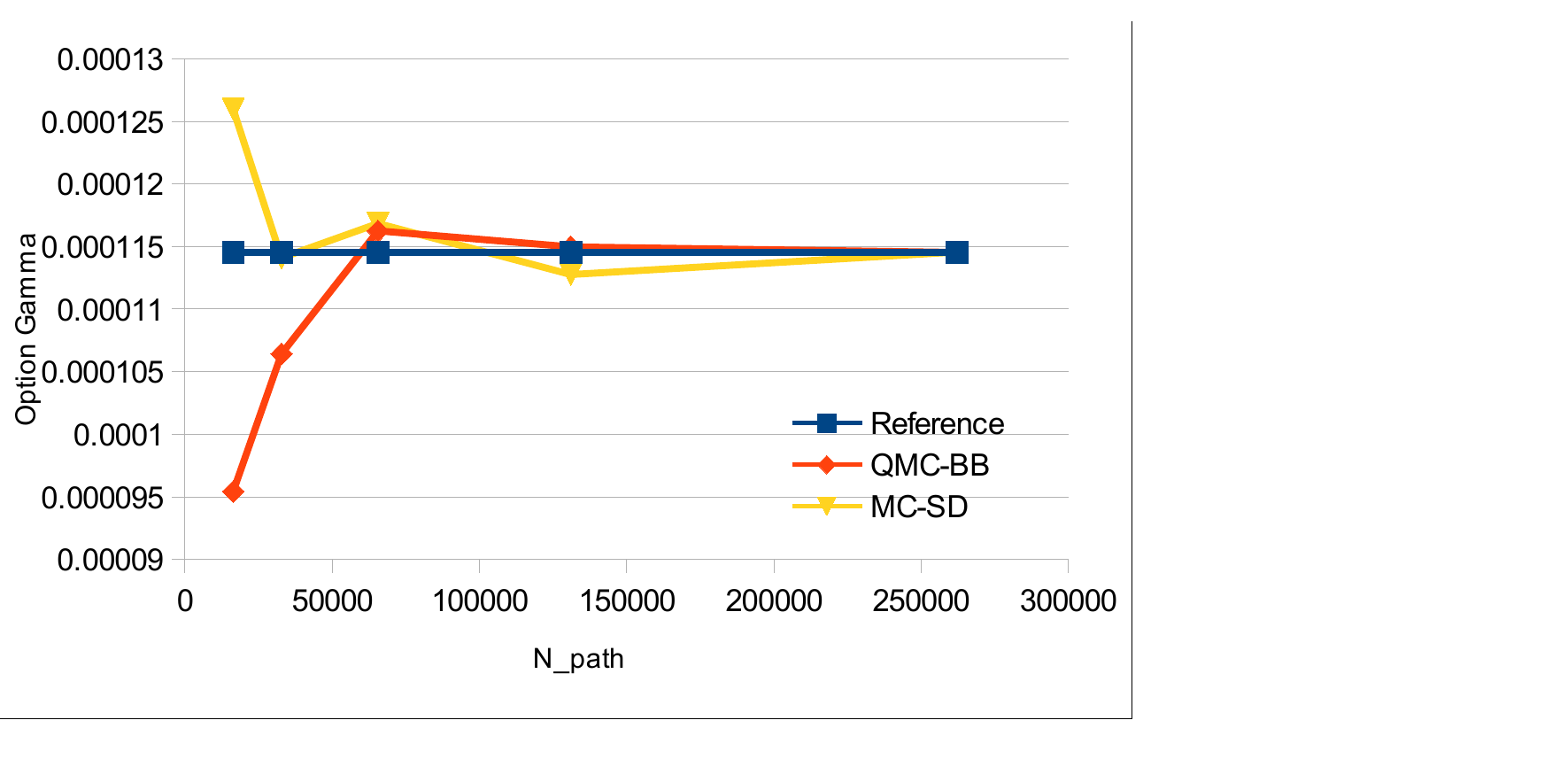}
    \label{figure:OptionGammaK80}
  \end{minipage}
  }
  \hfill
   \subfigure[]
{
  \begin{minipage}[b]{5.5cm}
    \includegraphics[width=8cm, height=4cm, angle =0,trim={2cm 18.5cm 1cm 2.5cm}, clip]{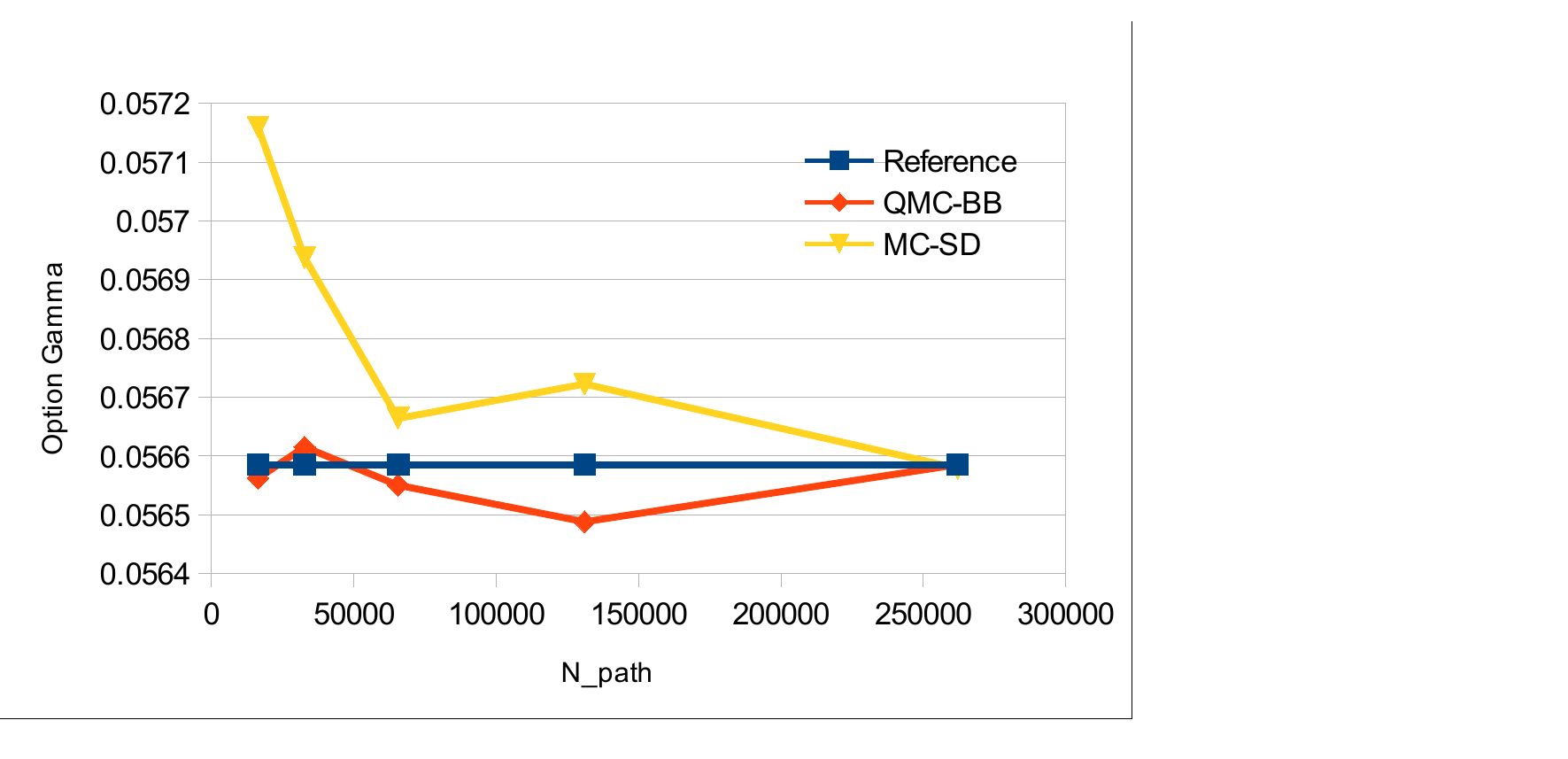}
    \label{figure:OptionGammaK100}
  \end{minipage}
  }
   \subfigure[]
{
   \begin{minipage}[b]{5.5cm}
    \includegraphics[width=8cm, height=4cm, angle =0,trim={2cm 18.5cm 1cm 2.4cm}, clip]{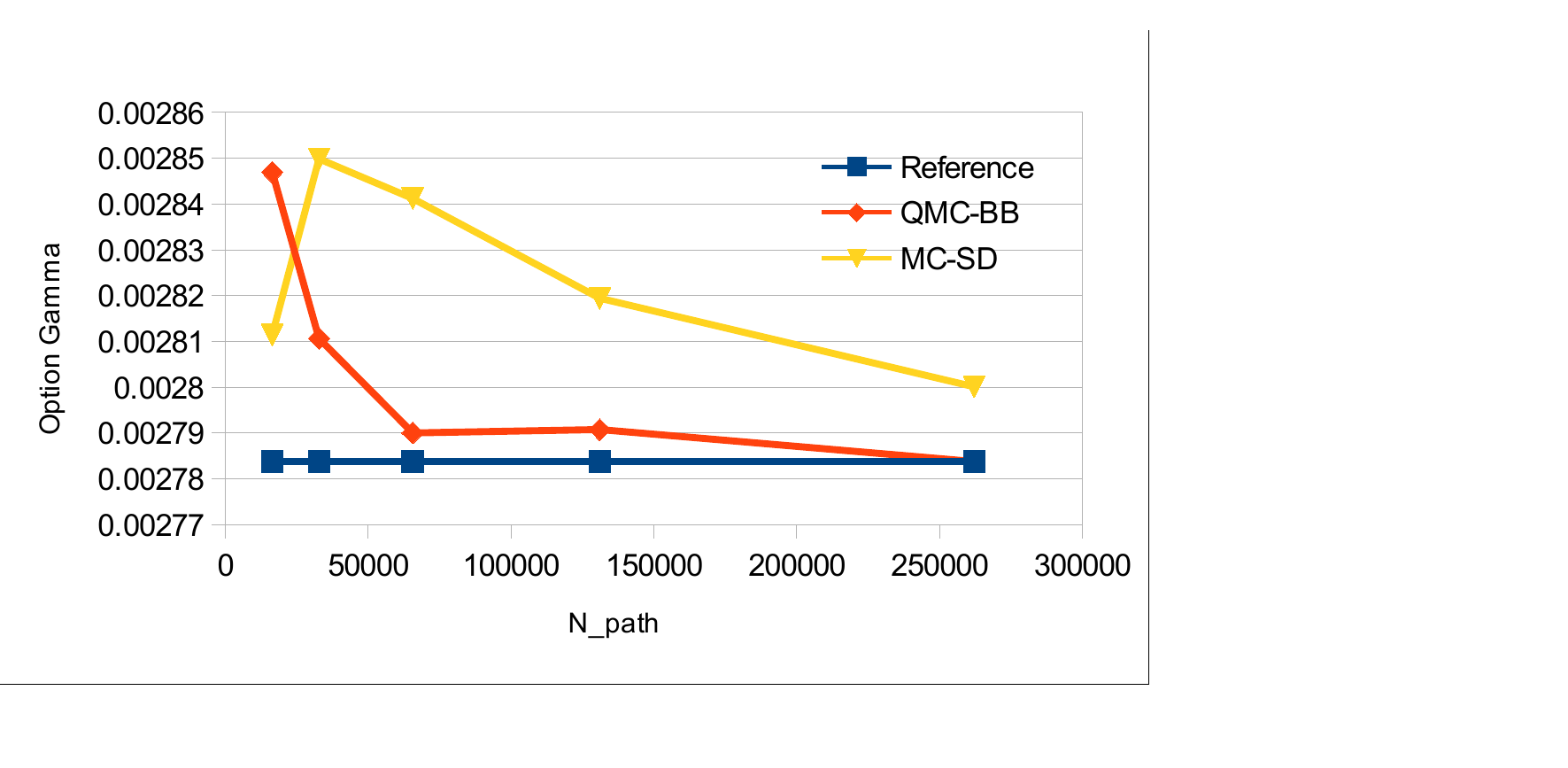}
		\label{figure:OptionGammaK120}
  \end{minipage}
  }
\caption{Asian call gamma value with strike (a) 80; (b) 100; (c) 120.}
\label{figure:OptionGammaK80-120}
\end{figure}

\begin{figure}[!tbp]
  \centering
\subfigure[]
{
  \begin{minipage}[b]{5.5cm}
    \includegraphics[width=8cm, height=4cm, angle =0,trim={2cm 18.5cm 1cm 2.5cm}, clip]{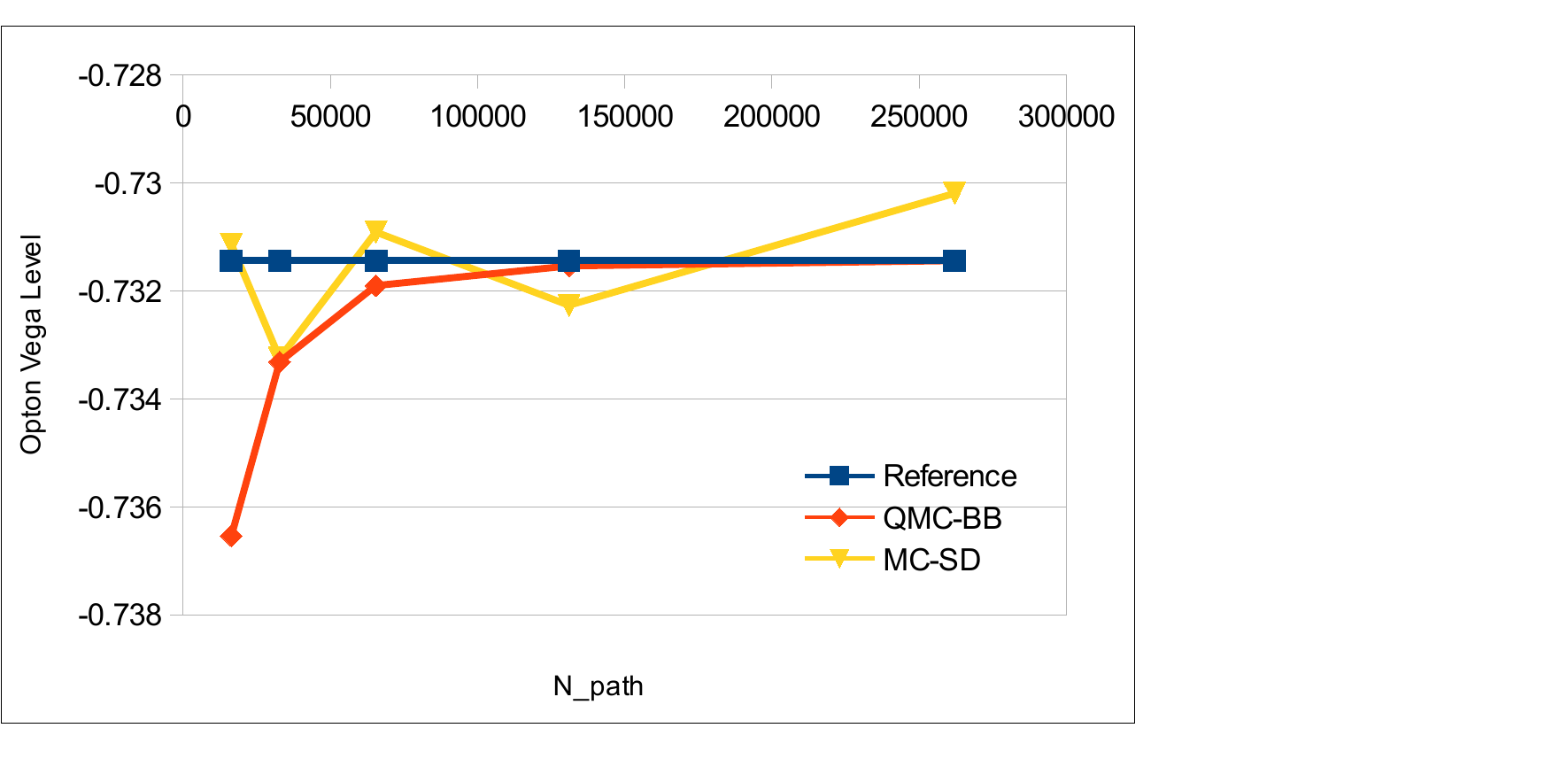}
    \label{figure:OptionVegaLevelK80}
  \end{minipage}
  }
  \hfill
  \subfigure[]
{
  \begin{minipage}[b]{5.5cm}
    \includegraphics[width=8cm, height=4cm, angle =0,trim={2cm 18.5cm 1cm 2.5cm}, clip]{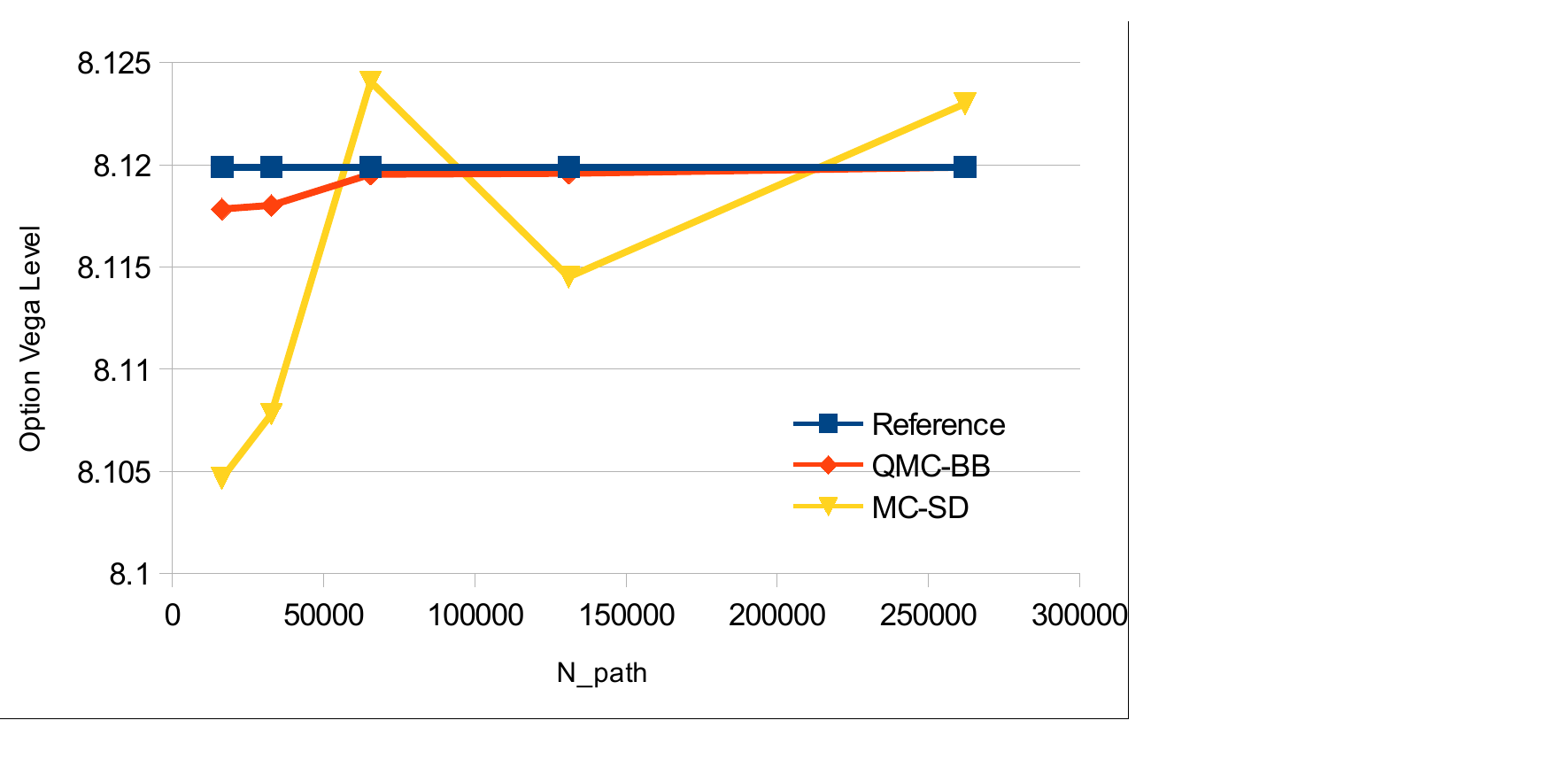}
    \label{figure:OptionVegaLevelK100}
  \end{minipage}
  }
  \subfigure[]
{
   \begin{minipage}[b]{5.5cm}
    \includegraphics[width=8cm, height=4cm, angle =0,trim={2cm 18.5cm 1cm 2.4cm}, clip]{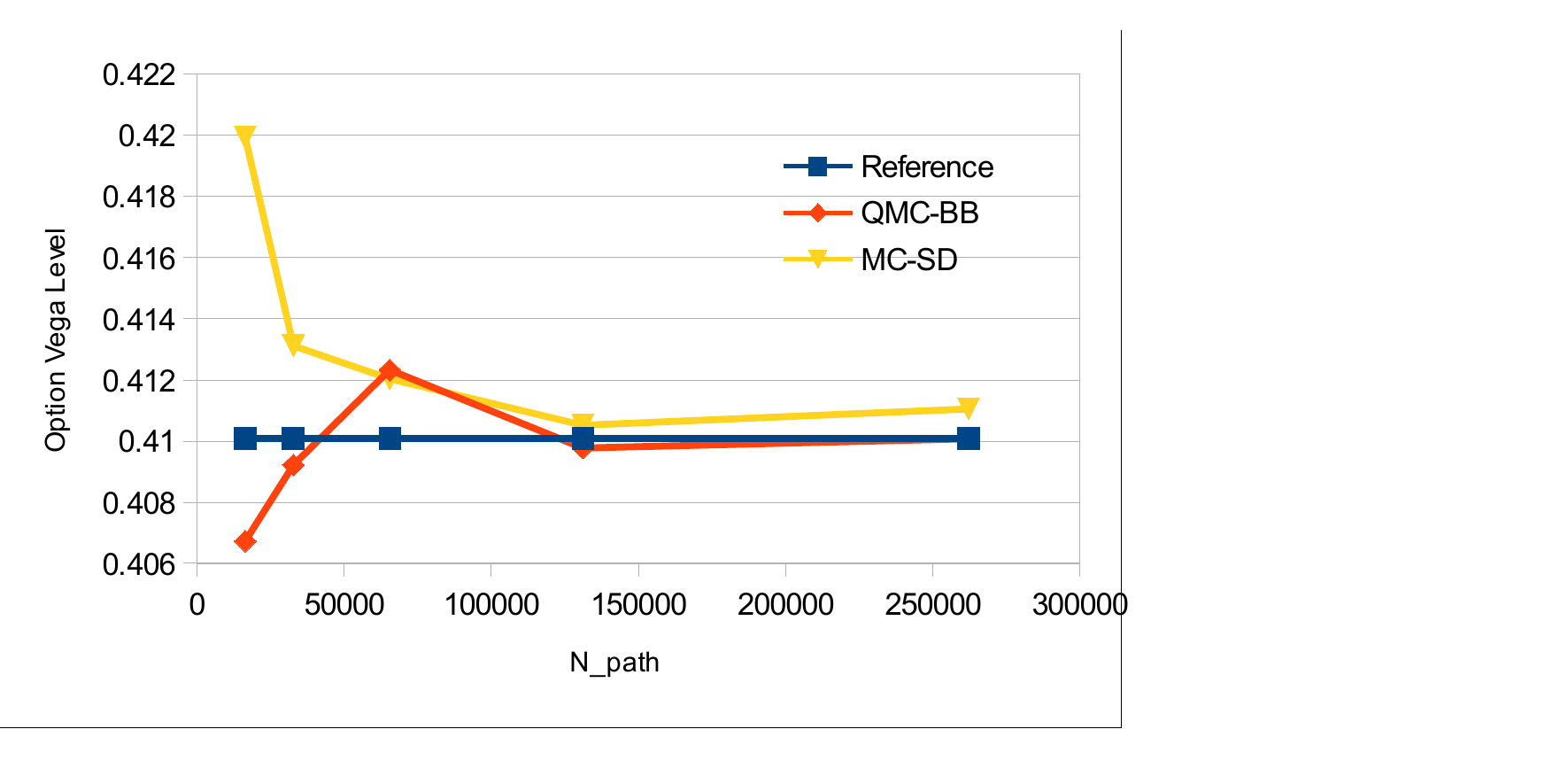}
		\label{figure:OptionVegaLevelK120}
  \end{minipage}
  }
 \caption{Asian call vega level value with strike (a) 80; (b) 100; (c) 120.}
 \label{figure:OptionVegaLevelK80-120}
\end{figure}

\begin{figure}[!tbp]
  \centering
  \subfigure[]
{
  \begin{minipage}[b]{5.5cm}
    \includegraphics[width=8cm, height=4cm, angle =0,trim={2cm 18.5cm 1cm 2.5cm}, clip]{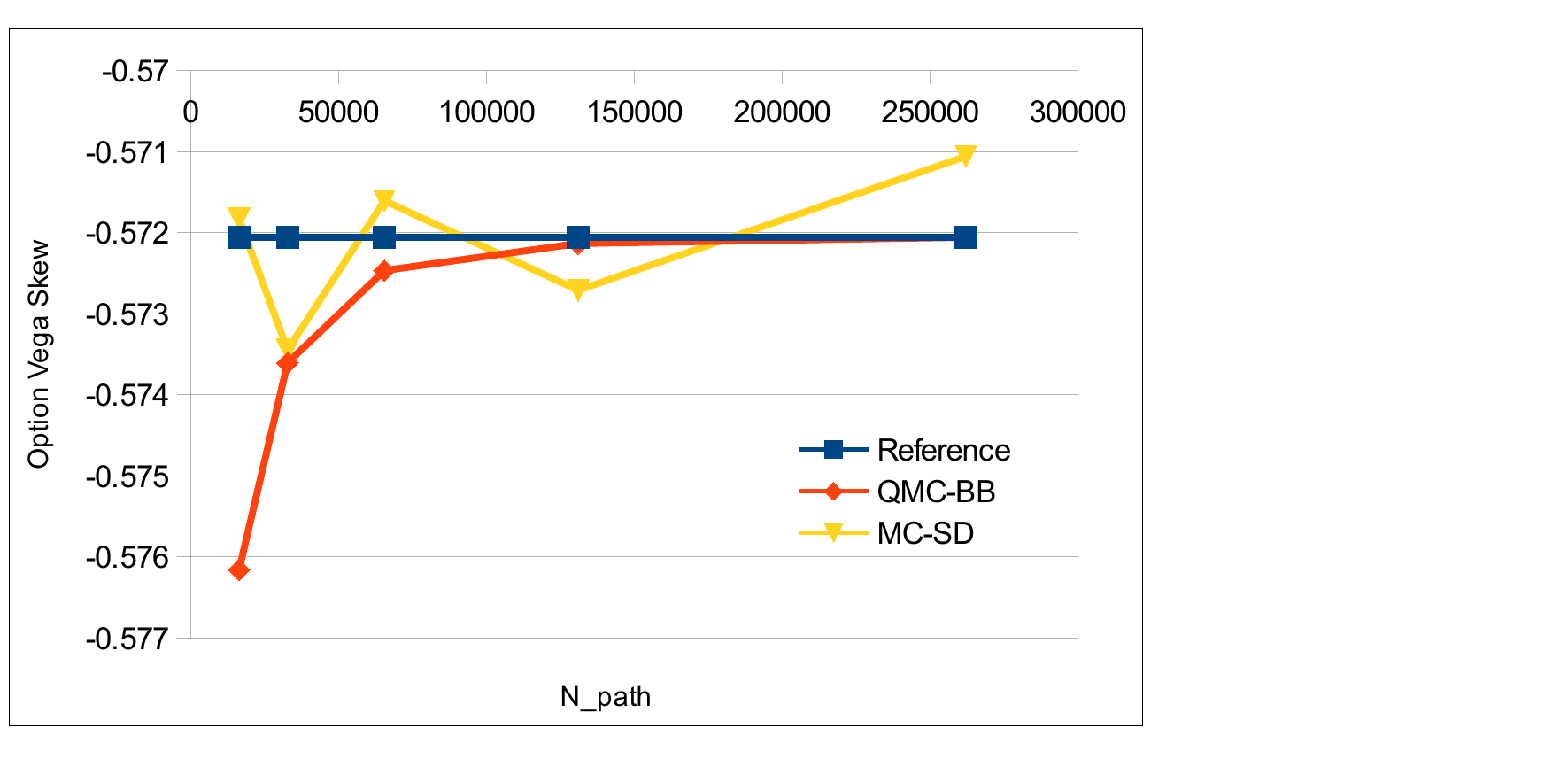}
    \label{figure:OptionVegaSkewK80}
  \end{minipage}
  }
  \hfill
  \subfigure[]
{
  \begin{minipage}[b]{5.5cm}
    \includegraphics[width=8cm, height=4cm, angle =0,trim={2cm 18.5cm 1cm 2.5cm}, clip]{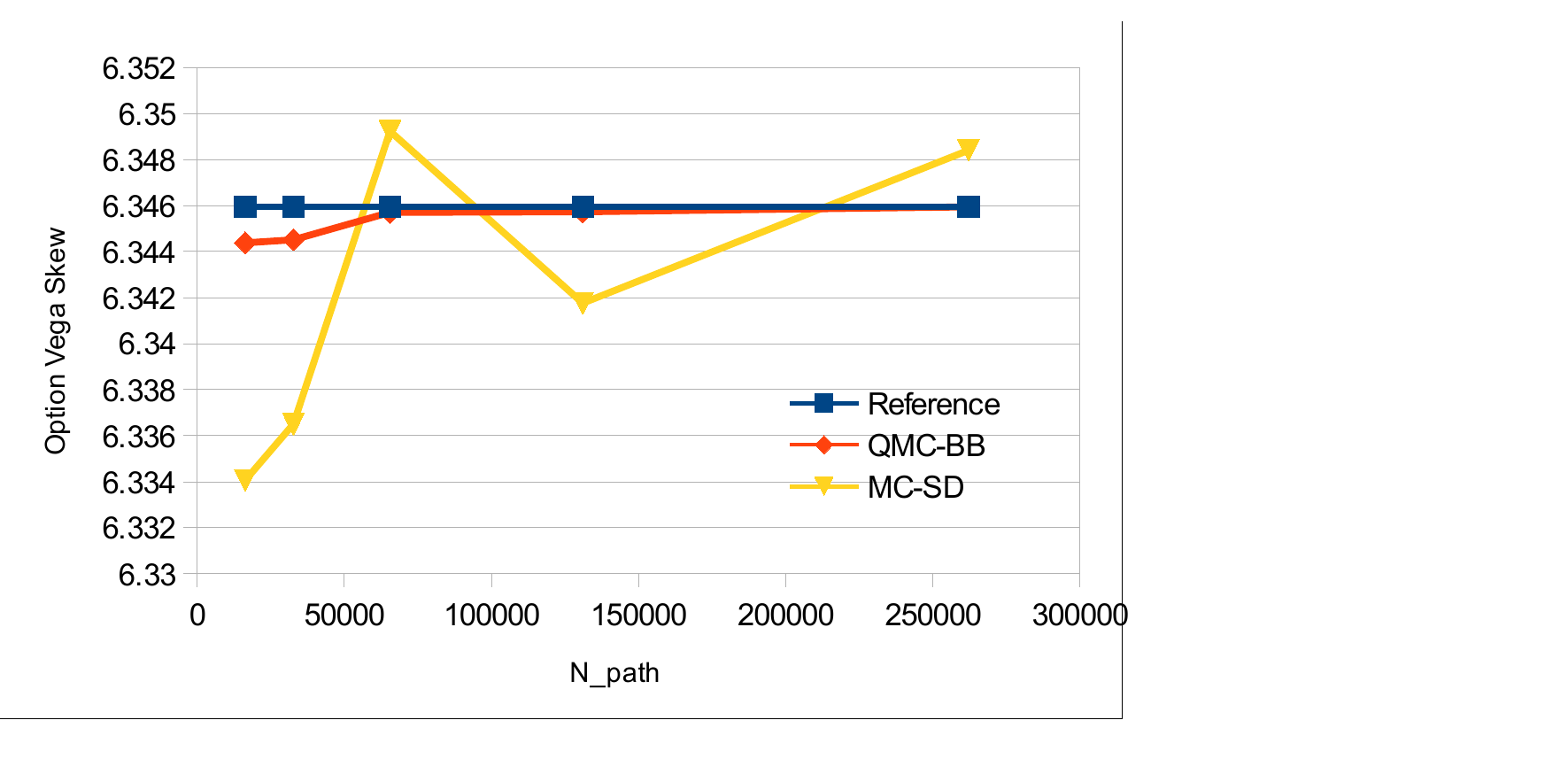}
    \label{figure:OptionVegaSkewK100}
  \end{minipage}
  }
  \subfigure[]
{
   \begin{minipage}[b]{5.5cm}
    \includegraphics[width=8cm, height=4cm, angle =0,trim={2cm 18.5cm 1cm 2.4cm}, clip]{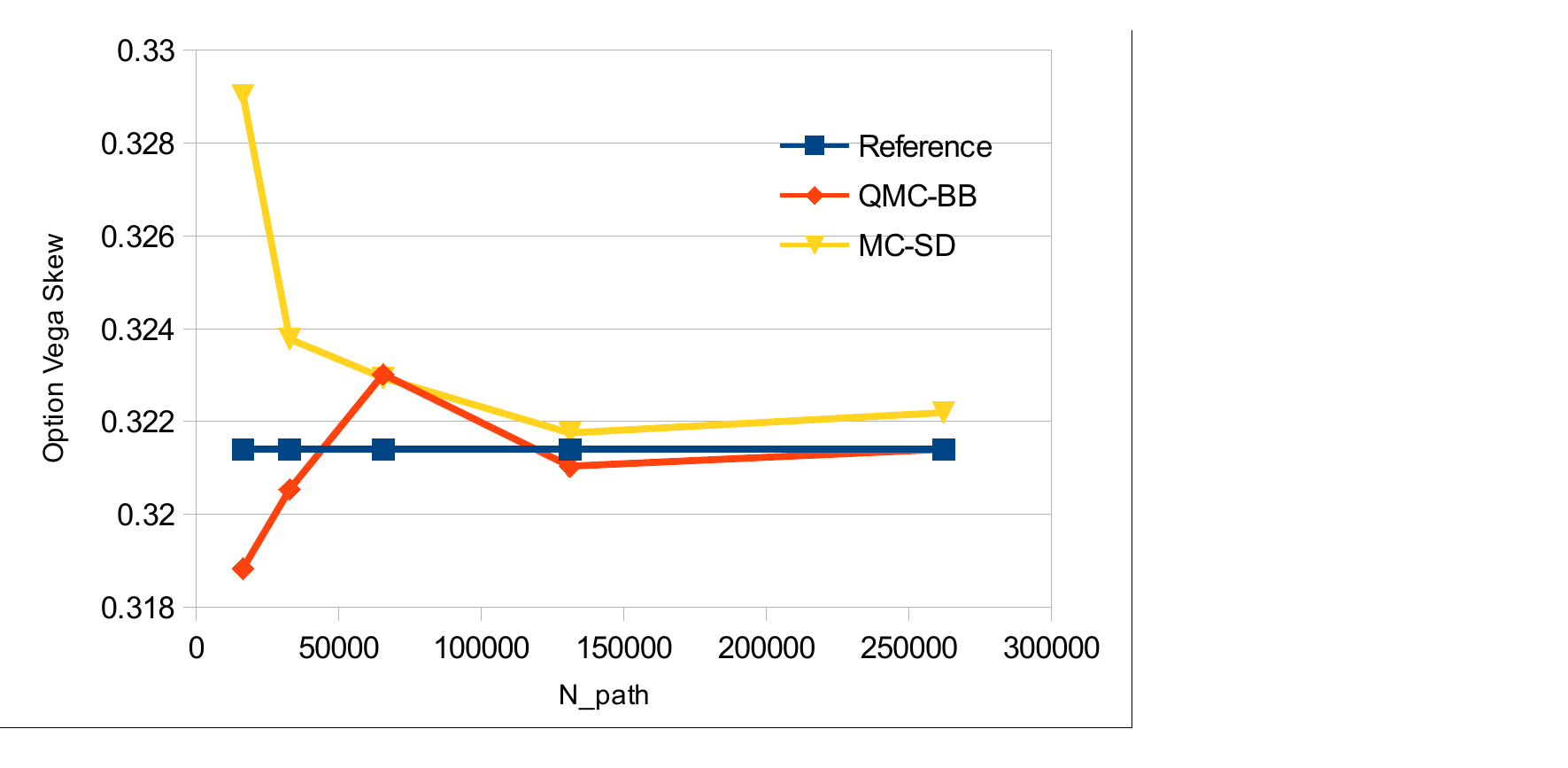}
		\label{figure:OptionVegaSkewK120}
  \end{minipage}
  }
    \caption{Asian call vega skew value with strike (a) 80; (b) 100; (c) 120.}
  	\label{figure:OptionVegaSkewK80-120}
\end{figure}

Figs. \ref{figure:OptionValueK80-120} to \ref{figure:OptionVegaSkewK80-120} show results of simulation of an Asian call price
and sensitivity factors versus the number of paths obtained using MC with the standard and QMC method with
the Brownian Bridge discretizations. We note that convergence of MC doesn't depend on the type of
discretization scheme, hence we did not show the results of MC with the Brownian Bridge discretization.
All results of MC simulation show that simulated solutions slowly converge to the reference solutions while the convergence curves are highly oscillating.
In contrast, QMC with Brownian Bridge is converging much faster and in a more stable way. It is also typically one-sided with a few exceptions.
We also notice slight variations in the speed of convergence depending whether option is
{\it{in-the-money}}, {\it{at-the-money}} or {\it{out-the-money}}. More efficient convergence behavior changes
depending on the type of the financial product: price or Greeks and it is different for different Greeks. For example for QMC for the call price
the most efficient ( fastest ) convergence is observed for {\it{at-the-money}} or {\it{out-the-money}} calls, while for delta - it is
{\it{in-the-money}} and {\it{at-the-money}}.

We also computed the root mean square error (RMSE) for each quantity (price and sensitivity factors) defined as

\begin{equation}
\epsilon = \left( \frac{1}{L} \sum_{l=1}^L (\bar{Q}_{ref} - Q_N^{(l)})^2 \right)^{\frac{1}{2}}.
\end{equation}

\begin{figure}[!tbp]
  \centering
  \subfigure[]
{
  \begin{minipage}[b]{5.7cm}
    \includegraphics[width=9cm, height=5cm, angle =0,trim={2cm 18.5cm 1cm 2.5cm}, clip]{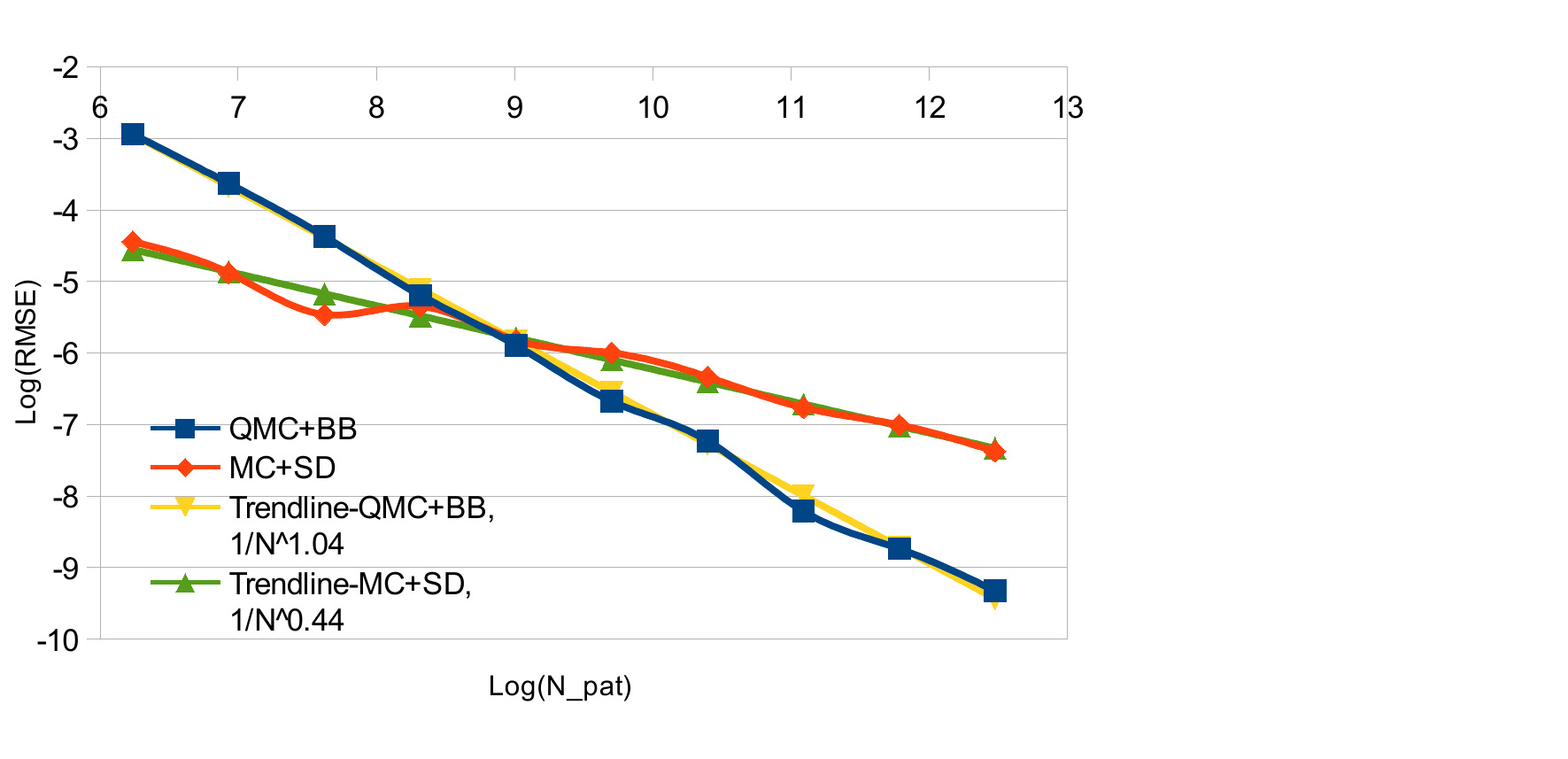}
   \label{figure:OptionValueK80CVRate}
  \end{minipage}
  }
  \hfill
  \subfigure[]
{
  \begin{minipage}[b]{5.7cm}
    \includegraphics[width=9cm, height=5cm, angle =0,trim={2cm 18.5cm 1cm 2.5cm}, clip]{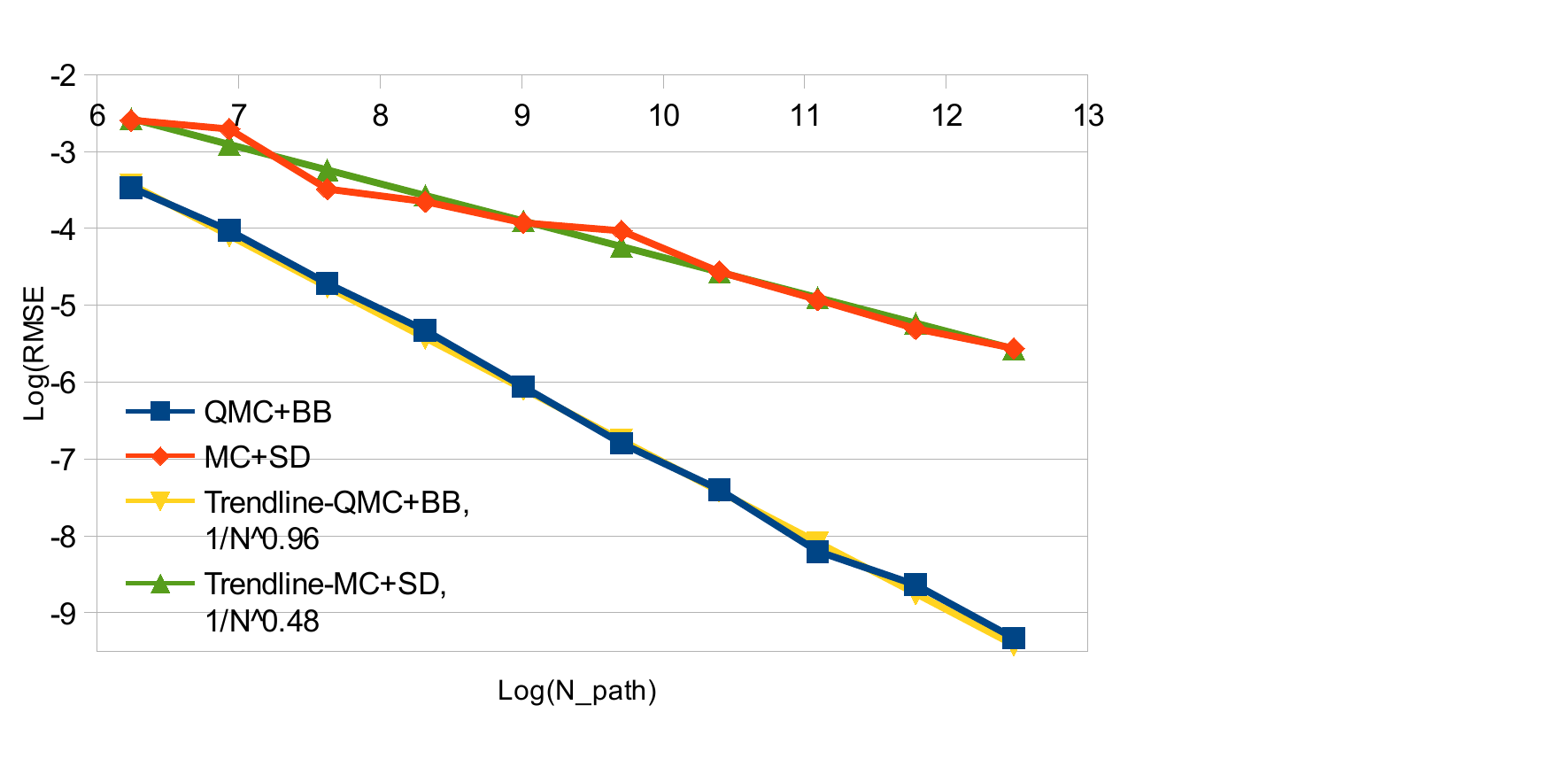}
    \label{figure:OptionValueK100CVRate}
  \end{minipage}
  }
\subfigure[]
{
   \begin{minipage}[b]{5.7cm}
    \includegraphics[width=9cm, height=5cm, angle =0,trim={2cm 18.5cm 1cm 2.4cm}, clip]{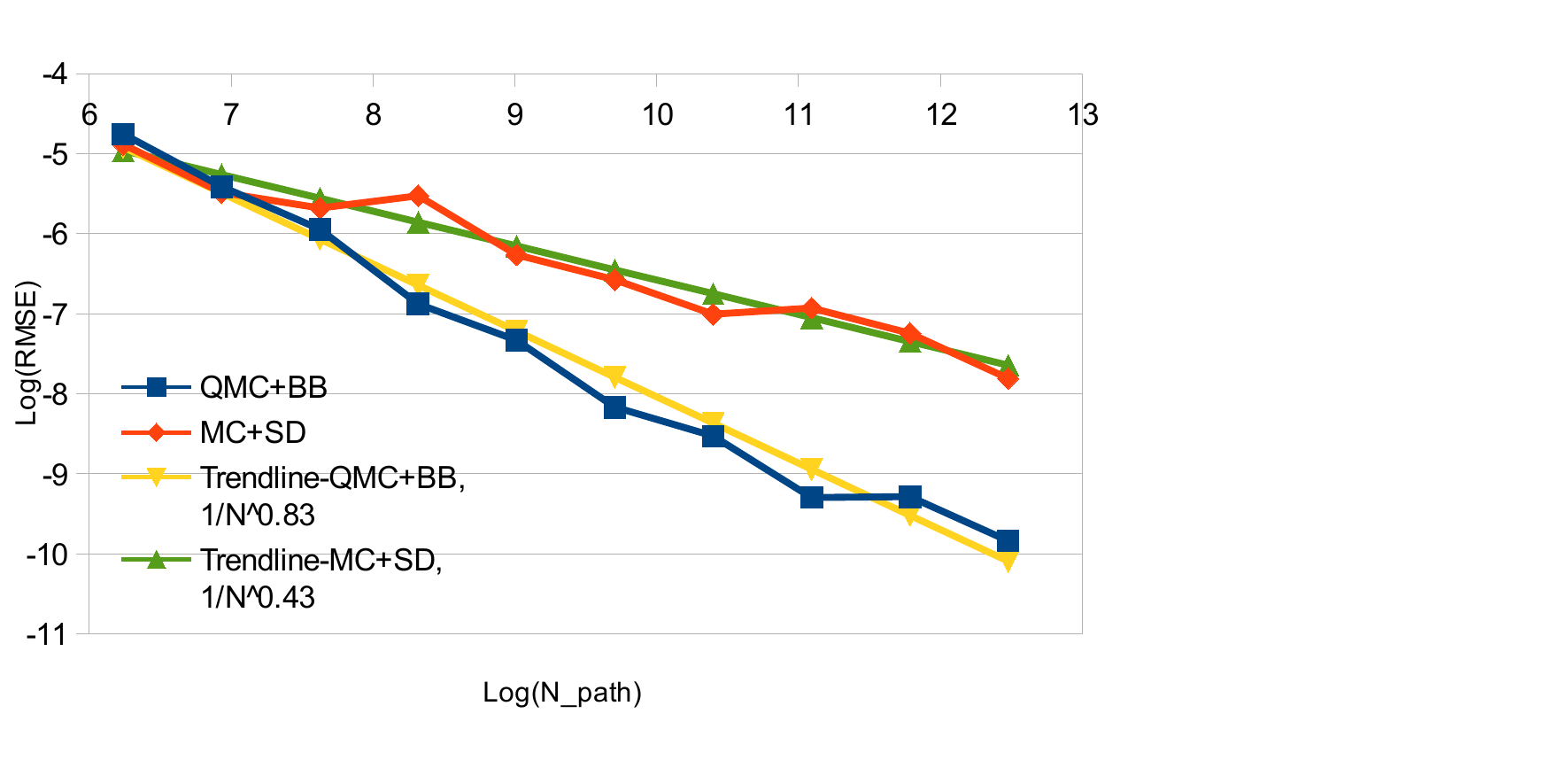}
 	\label{figure:OptionValueK120CVRate}
  \end{minipage}
  }
   \caption{Log-log plot of the root mean square error for Asian call price with strike (a) 80; (b) 100; (c) 120.}
  \label{figure:OptionValueK80-120CVRate}
\end{figure}

\begin{figure}[!tbp]
  \centering
  \subfigure[]
{
  \begin{minipage}[b]{5.7cm}
    \includegraphics[width=9cm, height=5cm, angle =0,trim={2cm 18.5cm 1cm 2.5cm}, clip]{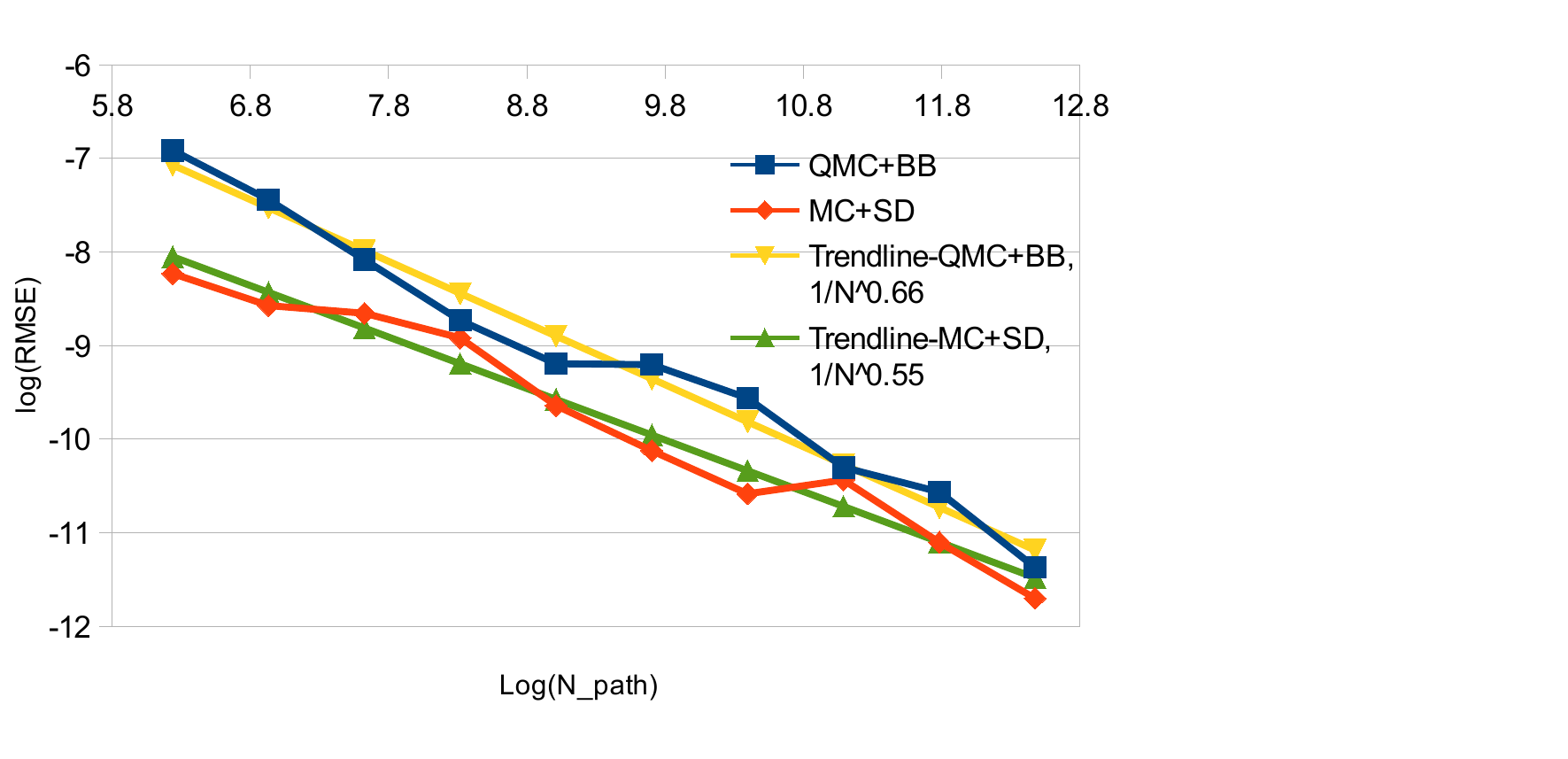}
    \label{figure:OptionDeltaK80CVRate}
  \end{minipage}
  }
  \hfill
  \subfigure[]
{
  \begin{minipage}[b]{5.7cm}
    \includegraphics[width=9cm, height=5cm, angle =0,trim={2cm 18.5cm 1cm 2.5cm}, clip]{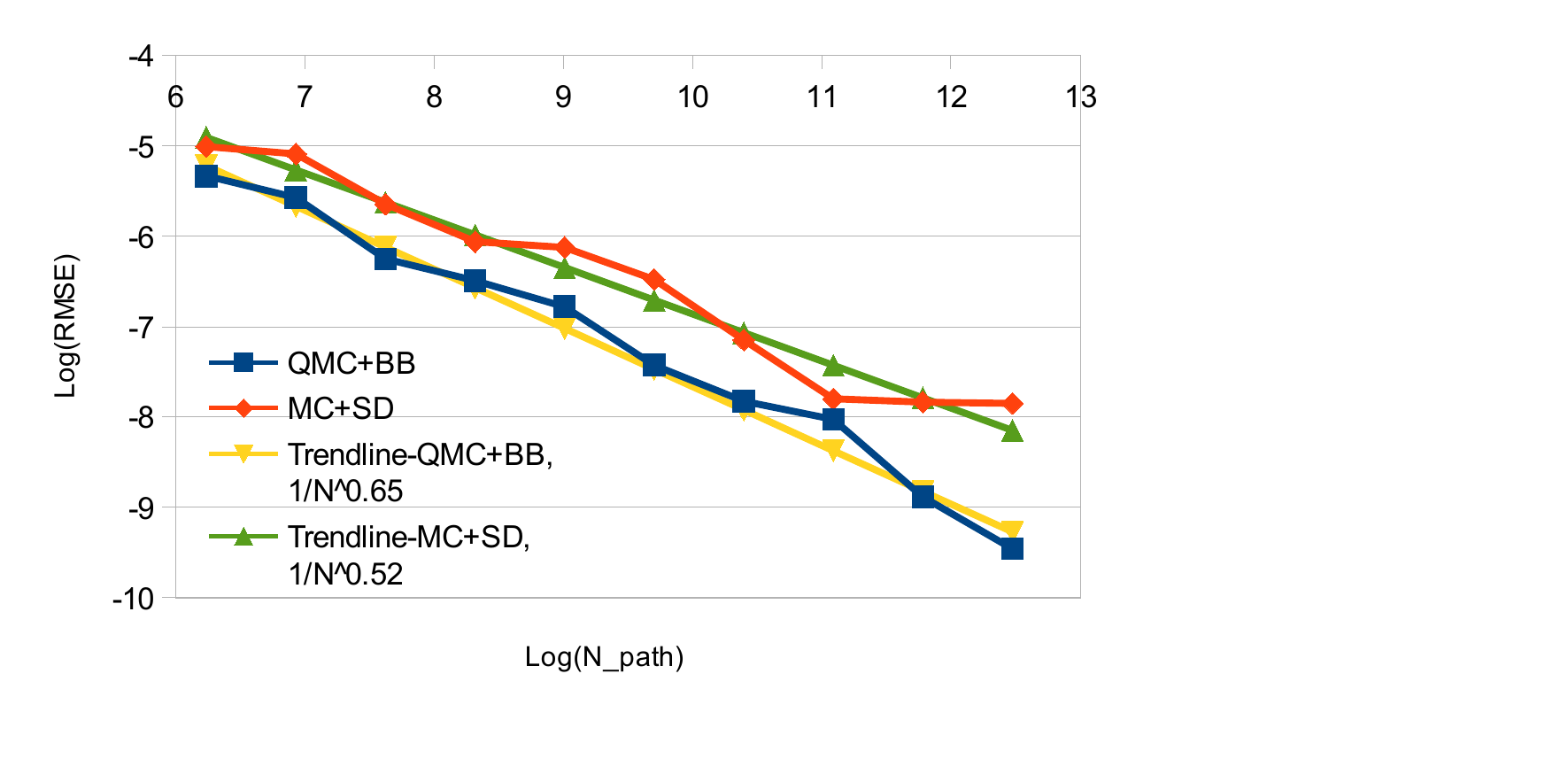}
    \label{figure:OptionDeltaK100CVRate}
  \end{minipage}
  }
  \subfigure[]
{
   \begin{minipage}[b]{5.7cm}
    \includegraphics[width=9cm, height=5cm, angle =0,trim={2cm 18.5cm 1cm 2.4cm}, clip]{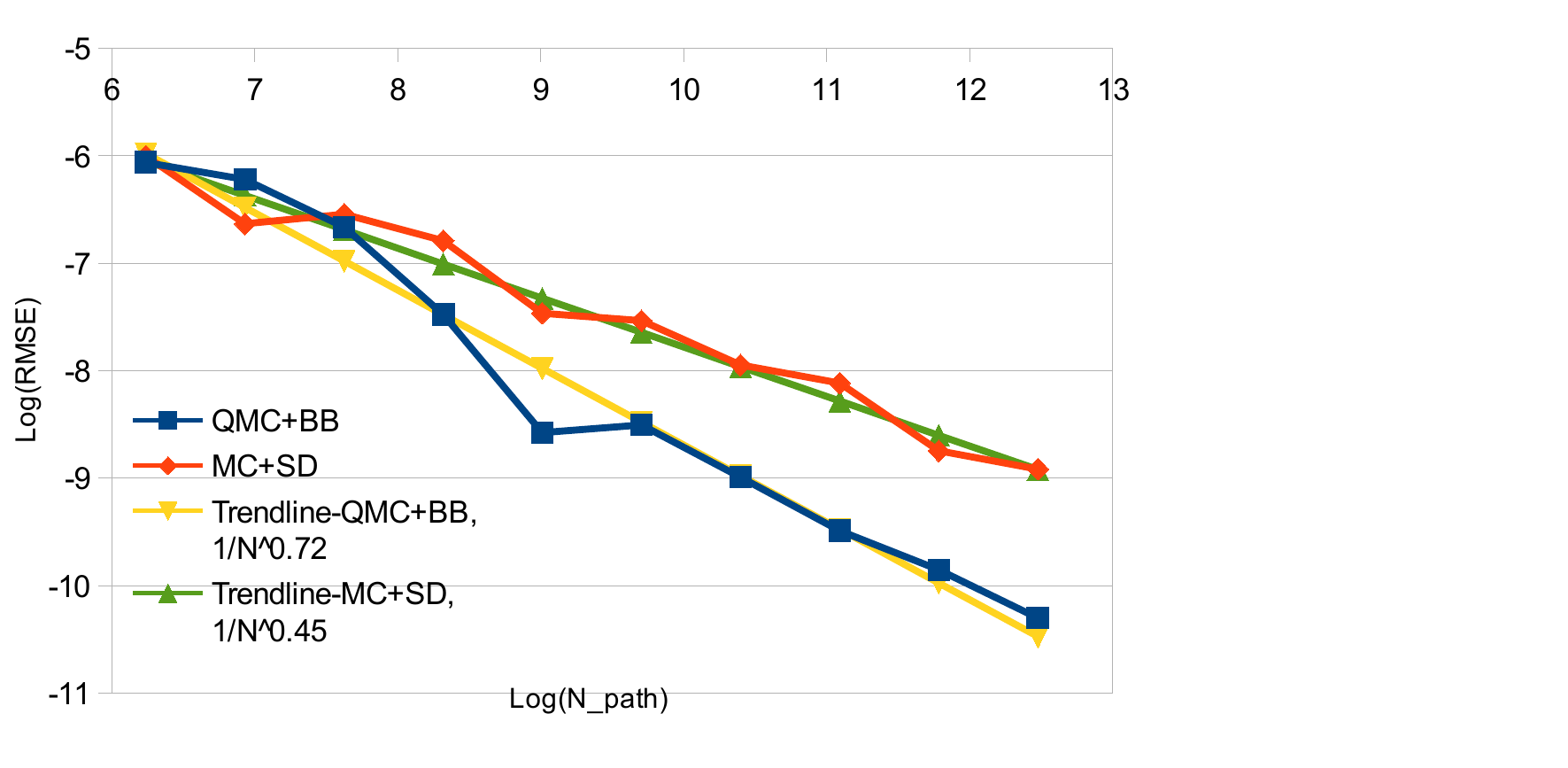}
		\label{figure:OptionDeltaK120CVRate}
  \end{minipage}
}
  \caption{Log-log plot of the root mean square error for Asian call delta with strike (a) 80; (b) 100; (c) 120.}
  \label{figure:OptionDeltaK80-120CVRate}
\end{figure}

\begin{figure}[!tbp]
  \centering
  \subfigure[]
{
  \begin{minipage}[b]{5.7cm}
    \includegraphics[width=9cm, height=5cm, angle =0,trim={2cm 18.5cm 1cm 2.5cm}, clip]{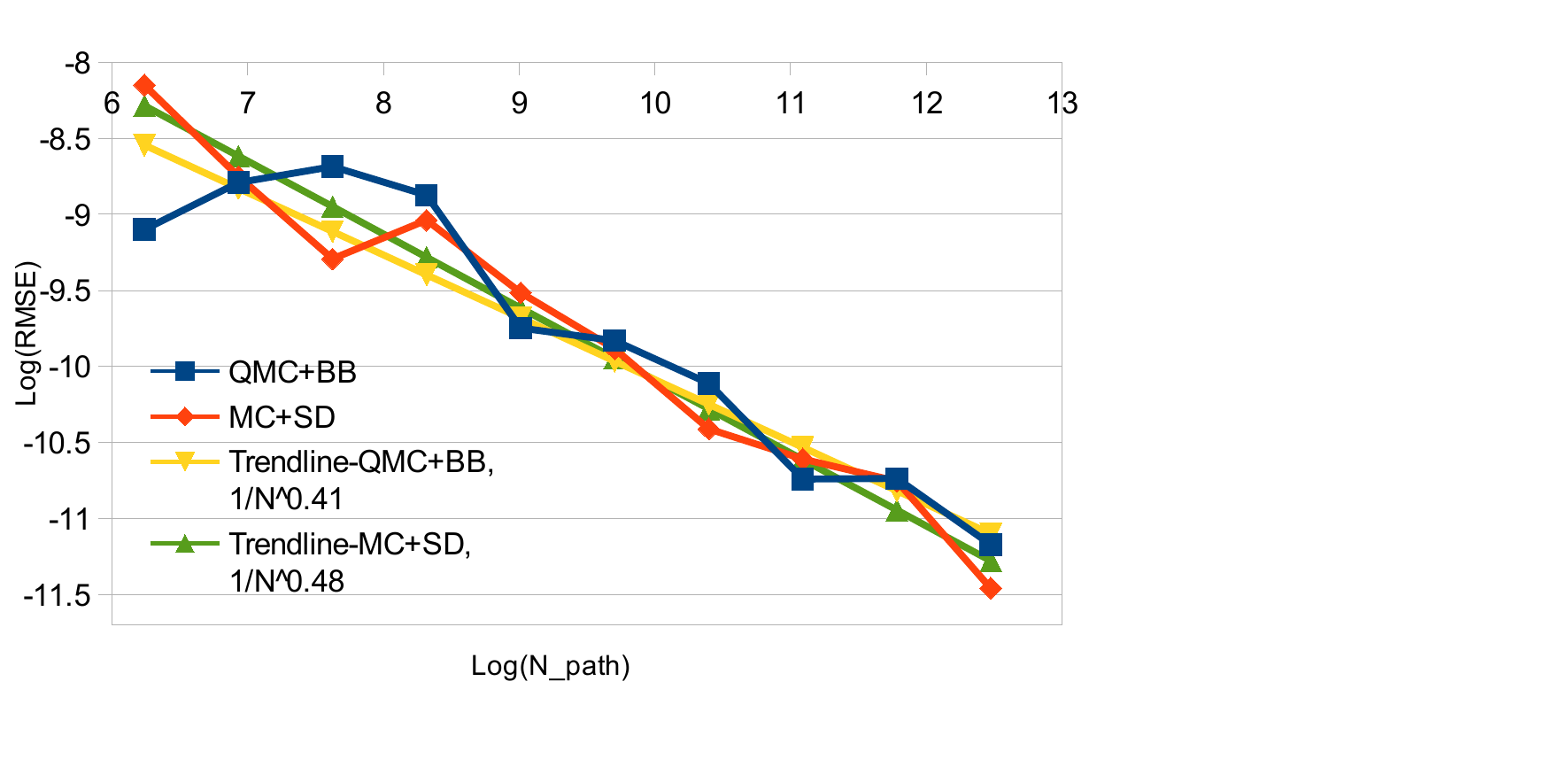}
    \label{figure:OptionGammaK80CVRate}
  \end{minipage}
  }
  \hfill
    \subfigure[]
{
  \begin{minipage}[b]{5.7cm}
    \includegraphics[width=9cm, height=5.3cm, angle =0,trim={2cm 18.9cm 1cm 2.5cm}, clip]{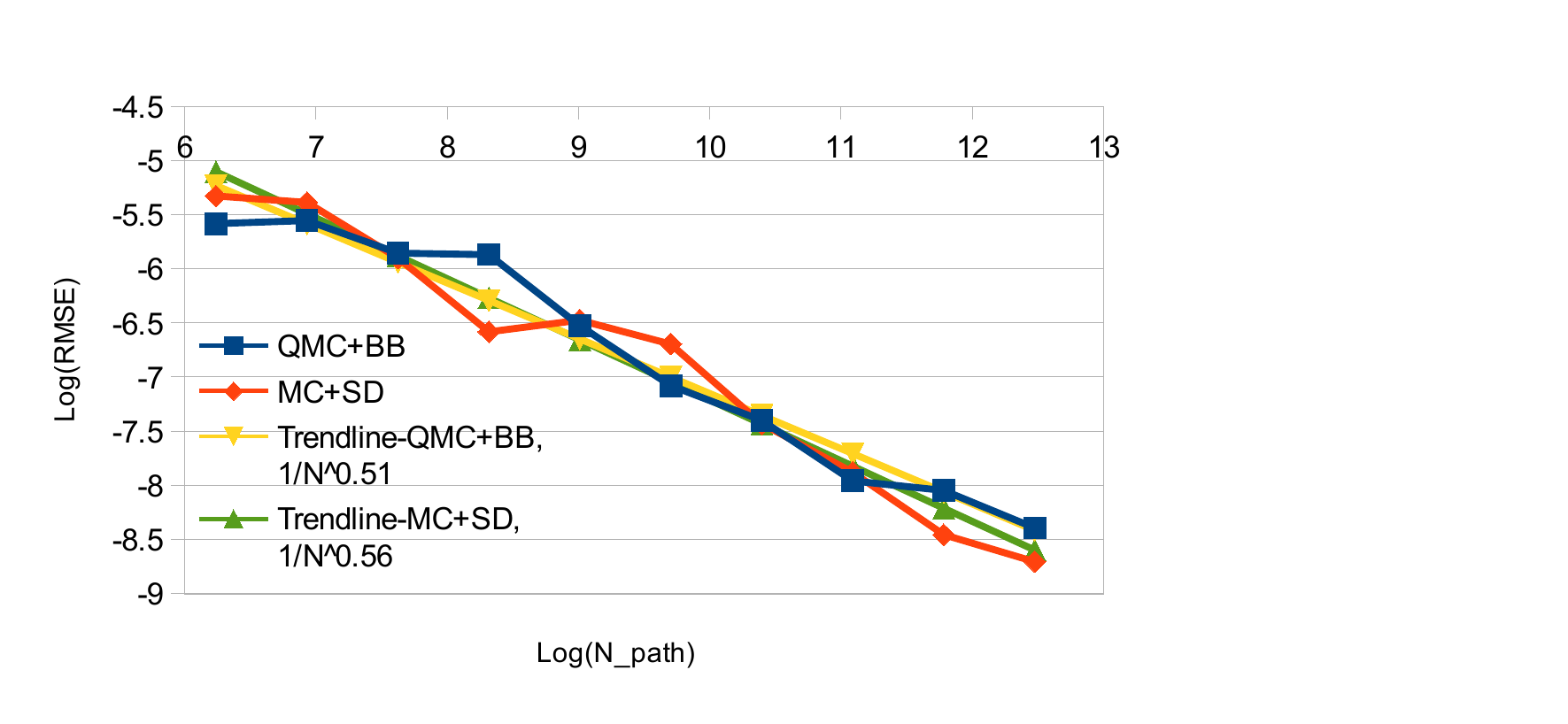}
    \label{figure:OptionGammaK100CVRate}
  \end{minipage}
  }
    \subfigure[]
{
   \begin{minipage}[b]{5.7cm}
    \includegraphics[width=9cm, height=5cm, angle =0,trim={2cm 18.5cm 1cm 2.4cm}, clip]{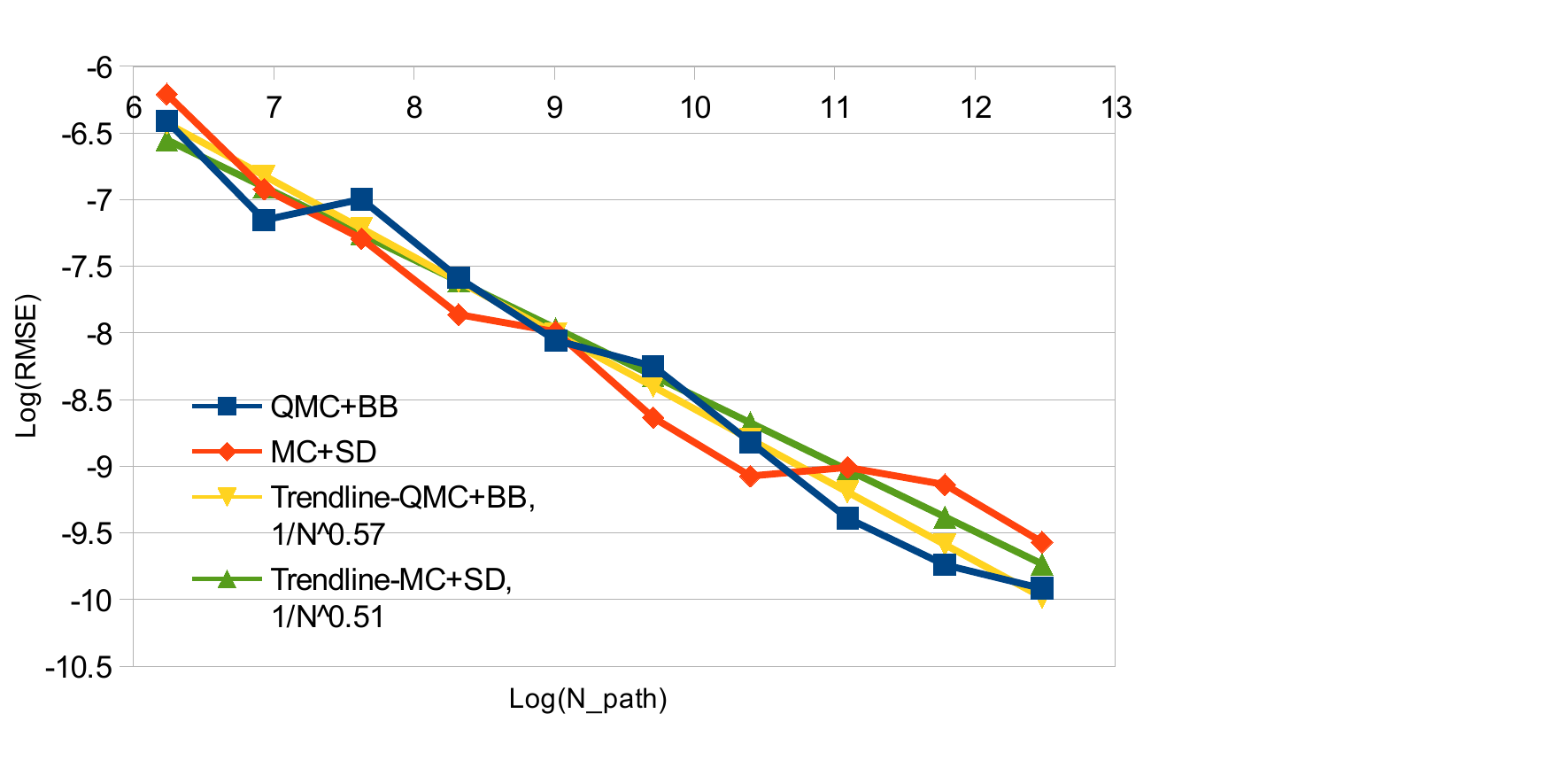}
		\label{figure:OptionGammaK120CVRate}
  \end{minipage}
  }
  \caption{Log-log plot of the root mean square error for Asian call gamma with strike (a) 80; (b) 100; (c) 120.}
  \label{figure:OptionGammaK80-120CVRate}
\end{figure}

\begin{figure}[!tbp]
  \centering
  \subfigure[]
{
  \begin{minipage}[b]{5.7cm}
    \includegraphics[width=9cm, height=5cm, angle =0,trim={2cm 18.5cm 1cm 2.5cm}, clip]{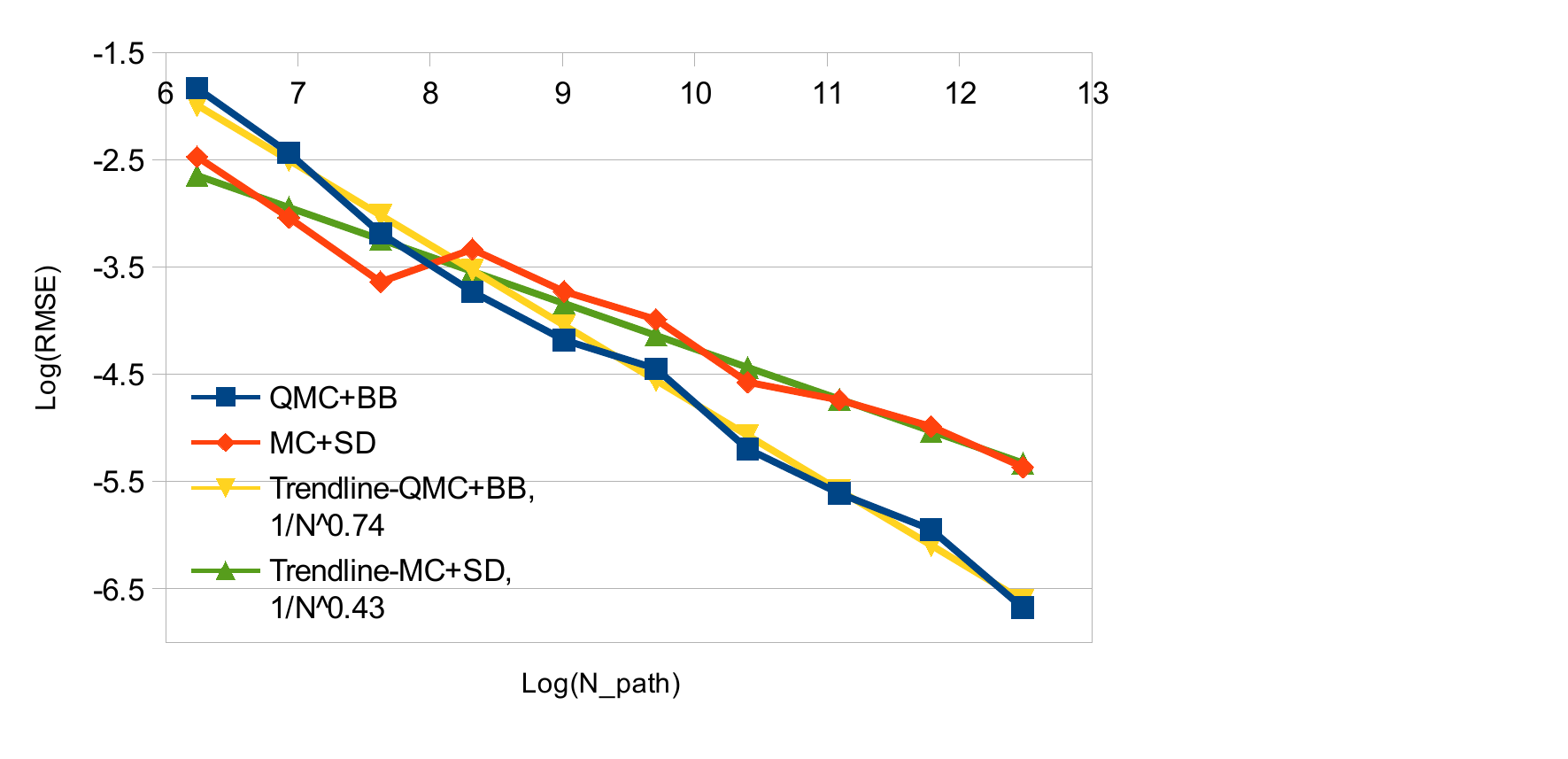}
    \label{figure:OptionVegaLevelK80CVRate}
  \end{minipage}
  }
  \hfill
  \subfigure[]
{
  \begin{minipage}[b]{5.7cm}
    \includegraphics[width=9cm, height=5cm, angle =0,trim={2cm 18.5cm 1cm 2.5cm}, clip]{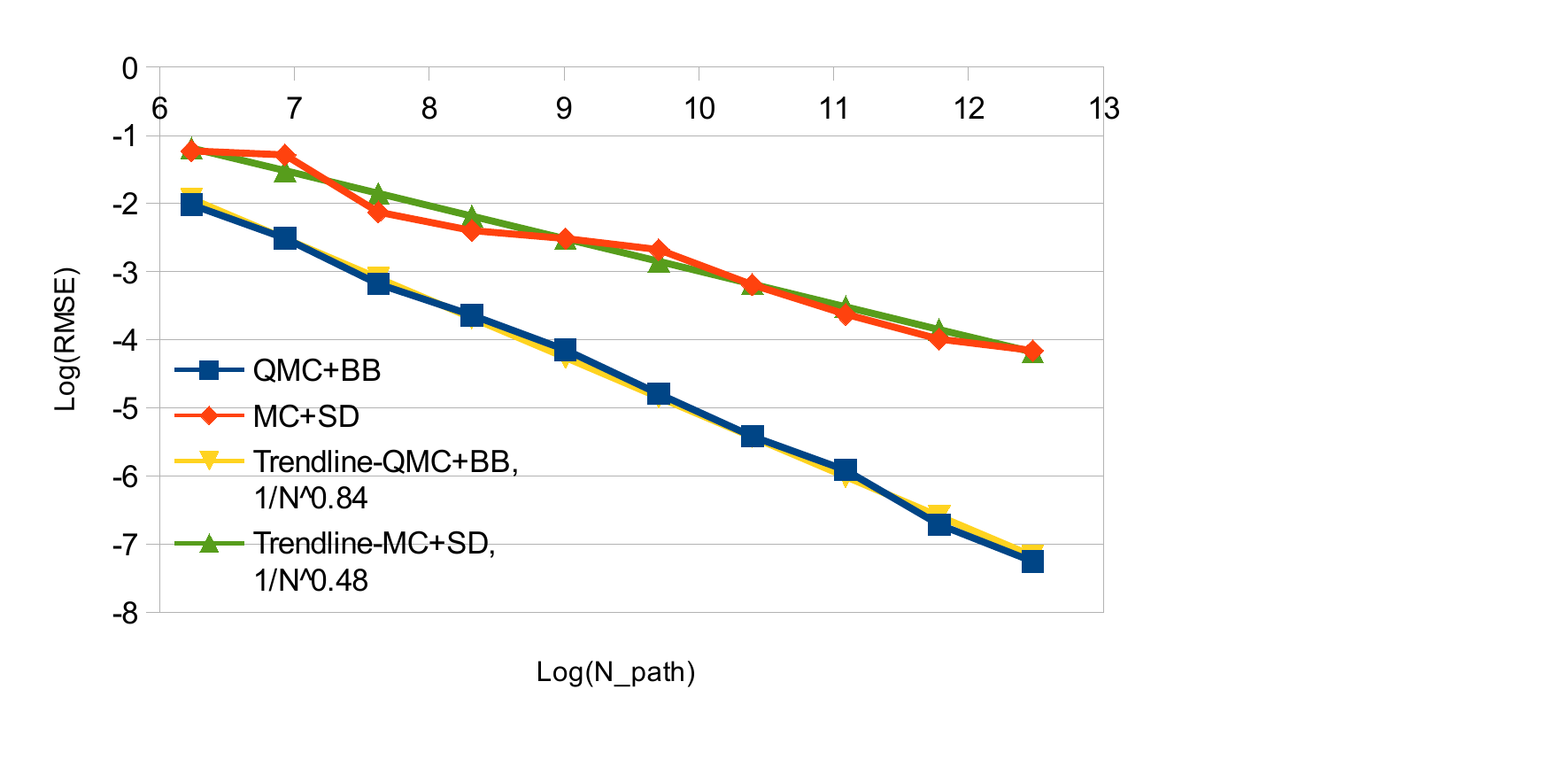}
    \label{figure:OptionVegaLevelK100CVRate}
  \end{minipage}
  }
  \subfigure[]
{
   \begin{minipage}[b]{5.7cm}
    \includegraphics[width=9cm, height=5cm, angle =0,trim={2cm 18.5cm 1cm 2.4cm}, clip]{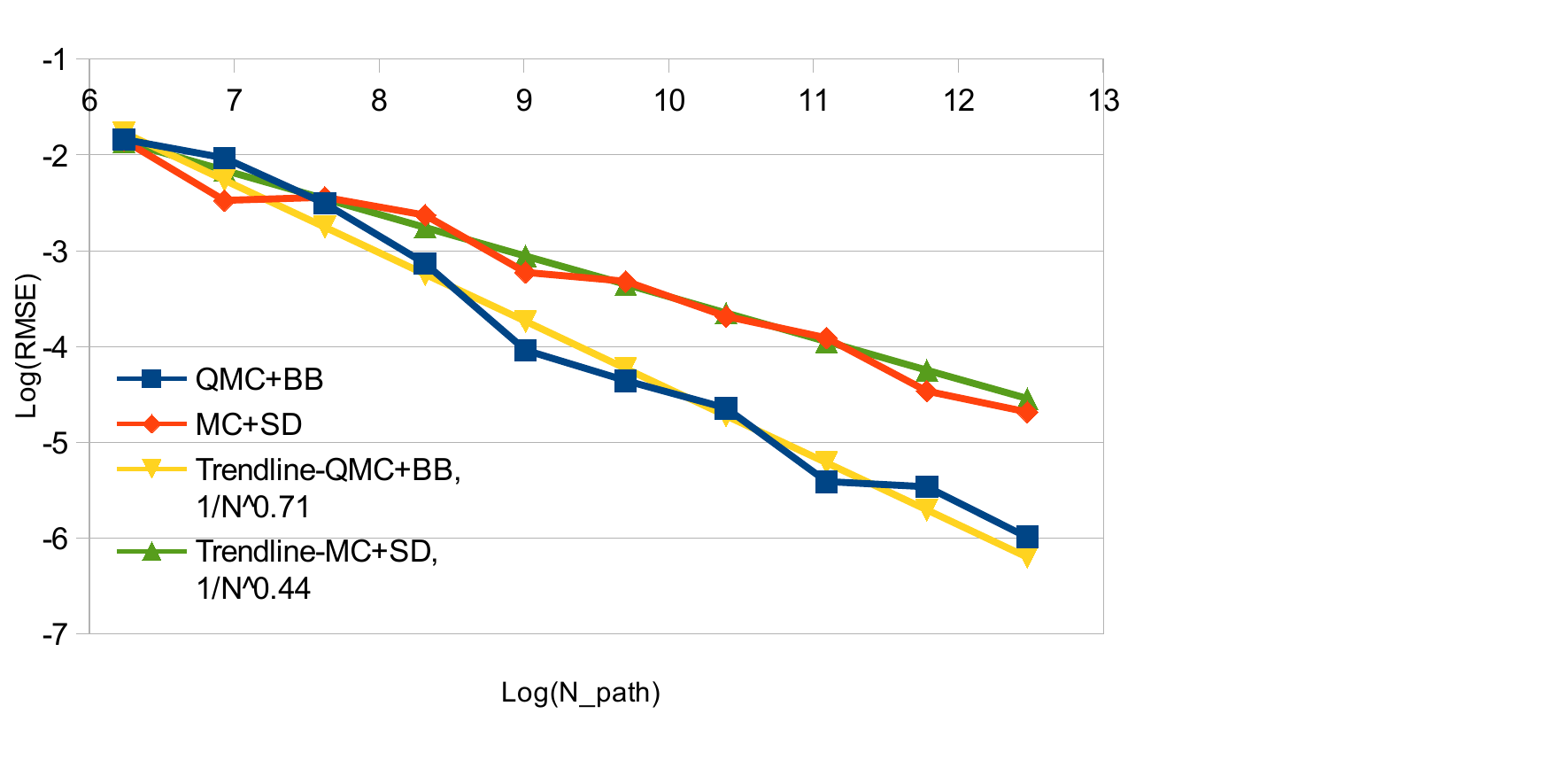}
		\label{figure:OptionVegaLevelK120CVRate}
  \end{minipage}
  }
   \caption{Log-log plot of the root mean square error for Asian call vega level with strike (a) 80; (b) 100; (c) 120.}
   \label{figure:OptionVegaLevelK80-120CVRate}
\end{figure}

\begin{figure}[!tbp]
  \centering
  \subfigure[]
{
  \begin{minipage}[b]{5.7cm}
    \includegraphics[width=9cm, height=5cm, angle =0,trim={2cm 18.5cm 1cm 2.5cm}, clip]{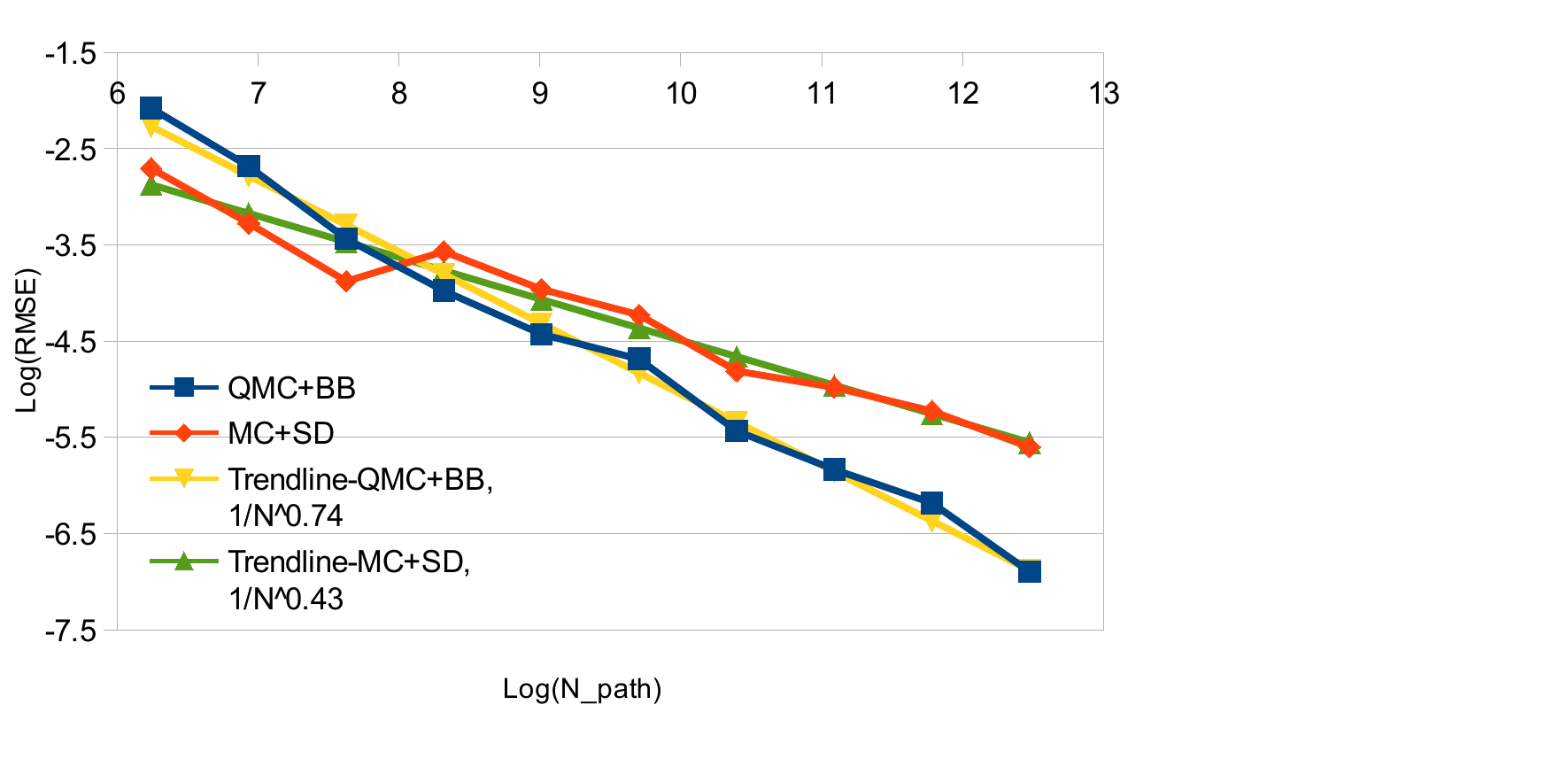}
    \label{figure:OptionVegaSkewK80CVRate}
  \end{minipage}
  }
  \hfill
  \subfigure[]
{
  \begin{minipage}[b]{5.7cm}
    \includegraphics[width=9cm, height=5cm, angle =0,trim={2cm 18.5cm 1cm 2.5cm}, clip]{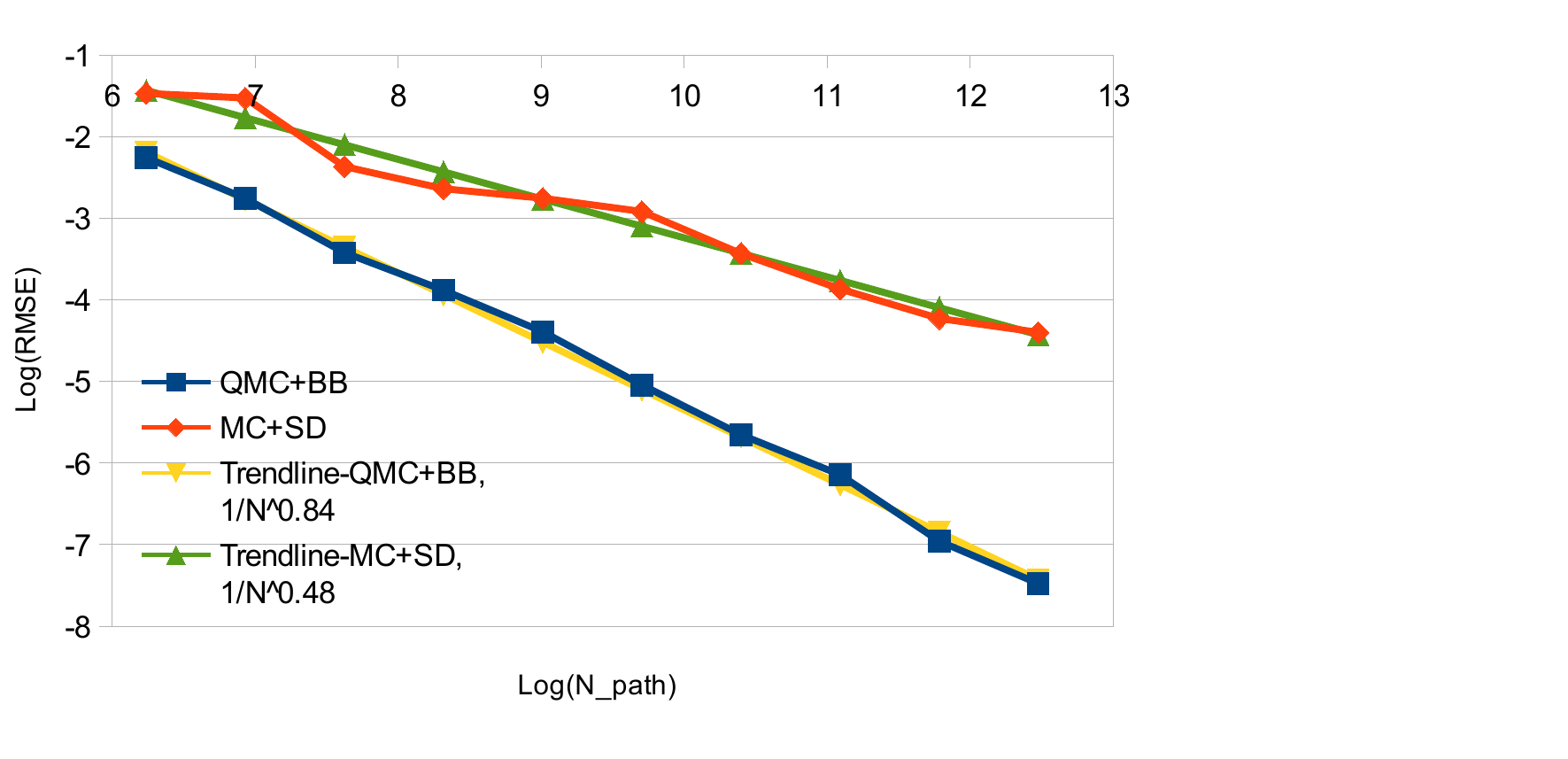}
    \label{figure:OptionVegaSkewK100CVRate}
  \end{minipage}
  }
  \subfigure[]
{
   \begin{minipage}[b]{5.7cm}
    \includegraphics[width=9cm, height=5cm, angle =0,trim={2cm 18.5cm 1cm 2.4cm}, clip]{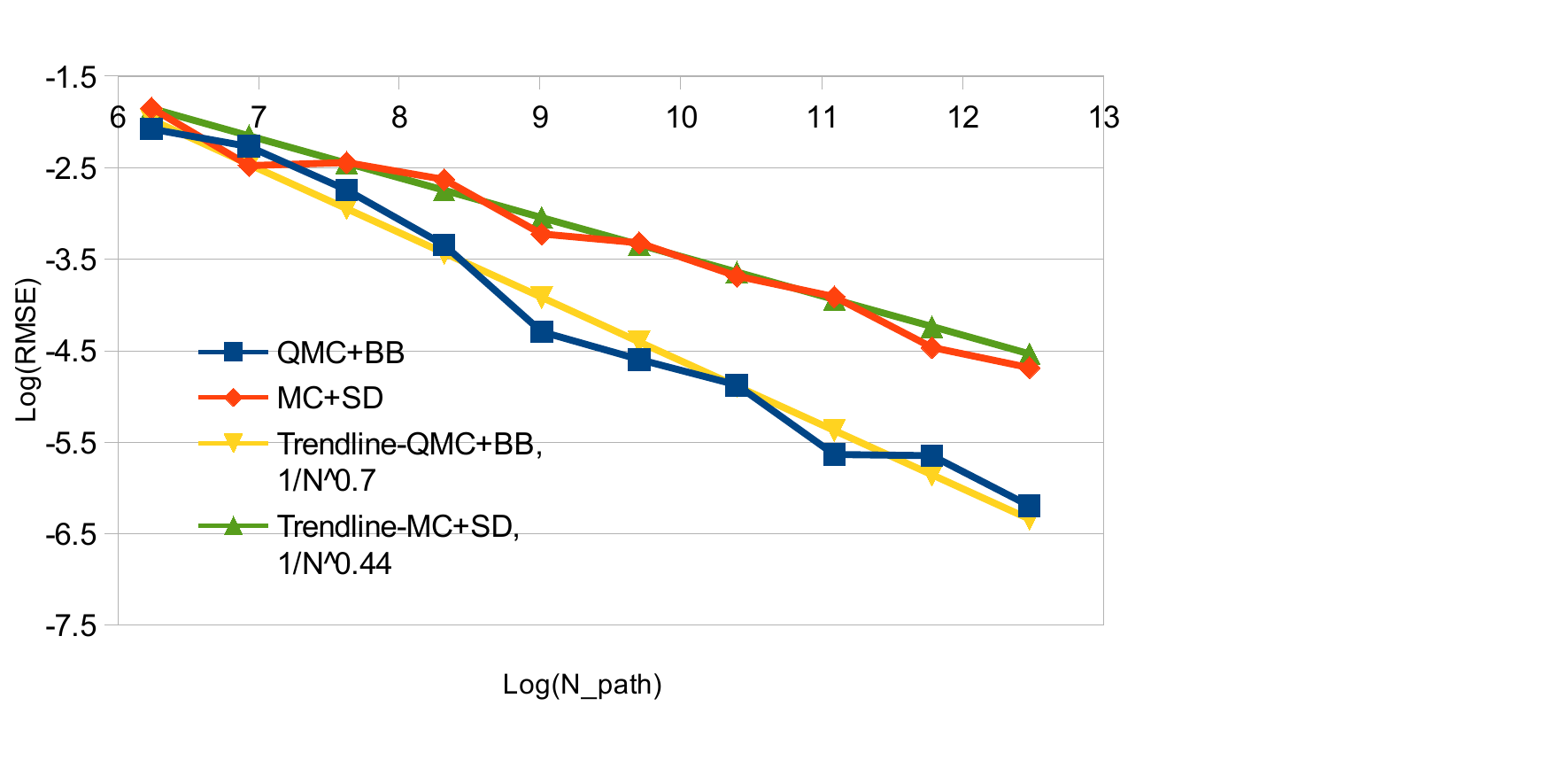}
		\label{figure:OptionVegaSkewK120CVRate}
  \end{minipage}
  }
  \caption{Log-log plot of the root mean square error for Asian call vega skew with strike (a) 80; (b) 100; (c) 120.}
		\label{figure:OptionVegaSkewK80-120CVRate}
\end{figure}

As in the previous computations, RMSE was averaged over $L=10$ independent runs. The Figs. \ref{figure:OptionValueK80-120CVRate}
to \ref{figure:OptionVegaSkewK80-120CVRate} show
the RMSE versus the number of paths for MC and QMC methods in a Log-Log scale.
We fitted the regression lines $N^{- \alpha}$
to extract convergence rates ${\alpha}$.
Depending on strike the convergence rate for QMC+BB
varies between 0.83 - 1.0 for Price,
0.65 - 0.72 for Delta,
0.5 for Gamma and 0.71 - 0.84 for Vega.
For the MC+SD this rate is very close to
the theoretically predicted for MC limit 0.5 (\ref{MCMeanSqError}).

\clearpage


\section{Conclusions}

We present and discuss the results of application
of MC and QMC methods for derivative pricing and risk analysis
based on Hyperbolic Local Volatility Model. Local volatility models usually capture the surface
of implied volatilities more accurately than other approaches.
The results presented for the Asian option show the superior performance
of the QMC methods especially for the Brownian Bridge
discretization scheme.
Effective dimensions fully explain the superior efficiency of QMC due to the specifics of Sobol' sequences.
The initial coordinates of Sobol’ LDS are much better distributed than the later
high dimensional coordinates. The Brownian Bridge
discretization scheme change the order in which time
steps are sampled. It uses well distributed coordinates from
each {\it{n}}-dimensional LDS vector to determine most of the structure of a path, and reserves other coordinates
to fill in finer details. This results in a reduction of the effective dimensions and
significantly improved accuracy of the QMC method.

\newpage

\bibliography{bibliography}
\bibliographystyle{plain}

\end{document}